\documentclass[apj]{emulateapj}
\usepackage{natbib}
\usepackage{float}
\begin{document}

\submitted{Accepted for publication in AJ}

\newcommand{\MITRGB}{\rm M_{I}(TRGB)}
\newcommand{\MI}{\rm M_{I}}

\shorttitle{Stellar Outskirts of M81}
\shortauthors{Barker et al.}

\title{Resolving the Stellar Outskirts of M81: \\
Evidence for a Faint, Extended
Structural Component
\footnote{Based on data collected
at Subaru telescope, which is operated by the 
National Astronomical Observatory of Japan.}}

\author{M. K. Barker and A. M. N. Ferguson}
\affil{Institute for Astronomy, University of Edinburgh, Blackford Hill, 
Edinburgh, EH9 3HJ, UK; mkb@roe.ac.uk}

\author{M. Irwin}
\affil{Institute of Astronomy, University of Cambridge, Madingley Road, 
Cambridge, CB3 0HA, UK}

\author{N. Arimoto}
\affil{(1) National Astronomical Observatory of Japan,
  Mitaka, Tokyo 181-8588, Japan              
   (2) Department of Astronomical Science, Graduate University for Advanced
Studies, Mitaka, Tokyo 181-8588, Japan}

\author{P. Jablonka}
\affil{Observatoire de Gen\'{e}ve, Laboratoire d'Astrophysique, Ecole Polytechnique F\'{e}d\'{e}rale de Lausanne (EPFL), CH-1290 Sauverny, Switzerland}

\begin{abstract} 

We present a wide field census of 
resolved stellar populations in 
the northern half of M81, 
conducted with Suprime-Cam on the 8-m Subaru telescope
and covering an area $\sim 0.3$ square degrees.  
The resulting color-magnitude diagram reaches over one magnitude below
the red giant branch (RGB) tip, allowing a detailed
comparison between the young and old stellar
spatial distributions.
The surface density of stars with ages $\lesssim 100$ Myr 
is correlated 
with that of neutral hydrogen in a manner similar to the
disk-averaged Kennicutt-Schmidt relation.
We trace this correlation 
down to gas densities of $\rm \sim 2 \times 10^{20}\ cm^{-2}$, 
lower than typically probed with $H\alpha$ flux.
Both diffuse light and resolved RGB star counts
show compelling evidence for a faint, extended structural component beyond
the bright optical disk, with a
much flatter surface brightness profile. 
The star counts 
allow us to probe this component to significantly 
fainter levels than is possible with the diffuse light alone.
From the colors of its RGB stars, 
we estimate this component has a peak global metallicity 
$\rm [M/H] \sim -1.1 \pm 0.3$ 
at deprojected radii 32 -- 44 kpc 
assuming an age of 10 Gyr and distance of 3.6 Mpc.
The spatial distribution of its RGB stars follows 
a power-law surface density profile, 
$I(r) \propto r^{-\gamma}$, with $\gamma \sim 2$.
If this component were separate from the bulge 
and from the bright optical disk, then it would contain 
$\sim 10 - 15\%$ of M81's total V-band luminosity.
We discuss the possibility that this is M81's halo
or thick disk, and in particular highlight its
similarities and differences with these components
in the Milky Way.
Other possibilities for its nature, 
such as a perturbed disk or
the faint extension of the bulge, cannot
be completely ruled out, 
though our data disfavor the latter. 
These observations add to the growing body of evidence
for faint, complex extended structures beyond the bright disks
of spiral galaxies.
\end{abstract}


\section{Introduction}

The stellar outskirts of spiral galaxies hold vital
clues to their formation and evolution.
Under the framework of the $\Lambda$CDM cosmology, 
the halos of massive spiral galaxies arise largely from the merging
and accretion of smaller sub-halos \citep{Bullock05}.
The stellar populations of spiral galaxy halos are, 
thus, directly linked to the properties of the sub-halos
and the accretion history \citep[e.g.,][]{Bullock05,Font06}.
Furthermore, the merging and accretion 
that is especially common in $\Lambda$CDM at high
redshift could be important in determining the properties
of thick disks, which can dominate the total disk light at
large radii and scale heights and 
which mounting evidence suggests are 
common in spiral galaxies 
\citep[e.g.,][]{Abadi03,Brook04,Dalcanton02,Yoachim06}.

Observing the outskirts of spiral galaxies is challenging
because these regions are very faint, typically at least several
magnitudes below the sky level, or $\mu_V \gtrsim 25 \rm\ mag\ arcsec^{-2}$.
Reliable surface brightness measurements at these faint levels 
require accurate flat-fielding, sky-determination, 
PSF modeling, and bright-star masking. 
Diffuse light studies of halos and thick disks are usually limited
to edge-on systems in which the thin disk light 
can be minimized by looking away from the midplane.
For example, by stacking the images of over 1000 edge-on galaxies
in the Sloan Digital Sky Survey (SDSS), \citet{Zibetti04} 
reached an $r$-band surface brightness $\mu_r \sim 31\ \rm mag\ arcsec^{-2}$.
They found a halo of excess emission around the stacked disk 
with an axis ratio of $c/a \sim 0.6$
and projected intensity distribution $I(r) \propto r^{-\gamma}$,
with $\gamma = 2$. 
In a related study, \citet{Zibetti04b} examined the extraplanar light
distribution around a nearly edge-on disk galaxy in the
Hubble Ultra Deep Field (HUDF) and found evidence for a halo
with similar structural properties as the SDSS stacked halo.
These structural properties are very similar to those of the
Milky Way (MW) halo \citep[][and references therein]{Bell08}, 
suggesting that such halos are common around disk galaxies.
However, the colors of the stacked SDSS and HUDF halos were 
much redder than expected for typical halo populations, 
suggesting an unusual stellar 
initial mass function (IMF), very high metallicity,
or significant scattered light from the disk 
\citep{Zibetti04,Zackrisson06,deJong08b}.

Studies of the diffuse light around galaxies have also
revealed evidence that thick disks are common
\citep[e.g.,][]{Burstein79,Tsikoudi79,Shaw90,deGrijs96,Morrison97,
Neeser02}.
\citet{Dalcanton02} obtained optical and near-infrared
images of many bulgeless edge-on disk galaxies reaching
$\mu_{R} \sim 27\ \rm mag\ arcsec^{-2}$.
From the color maps, vertical color gradients, and
faint isophote shapes, they
argued for the presence of red stellar envelopes around
all their galaxies with surface brightnesses, spatial distributions,
mean ages, and metallicities similar to the MW thick disk.
The uniformity in these properties is surprising
if they were assembled from satellites 
with different masses, gas fractions, histories, and
orbital properties, as expected in some hierarchical
structure formation models \citep[e.g.][]{Abadi03,Brook04}.
However, the ages and metallicities were very uncertain 
because of the unknown star formation histories, 
the shallow depth of the near-infrared images, and the 
inherent difficulties of sky-subtracting and
flat-fielding in the infrared.
Further analysis by \citet{Yoachim08,Yoachim08b} 
has indeed revealed a variety of thick disk kinematics and
compositions amongst the low-mass galaxies in the
\citet{Dalcanton02} sample.
The properties of the higher-mass galaxies' thick disks
were more difficult to constrain because their
thin disk light was too dominant in the regions with
adequate signal.

\begin{figure*}[t]
\epsscale{0.7}
\plotone{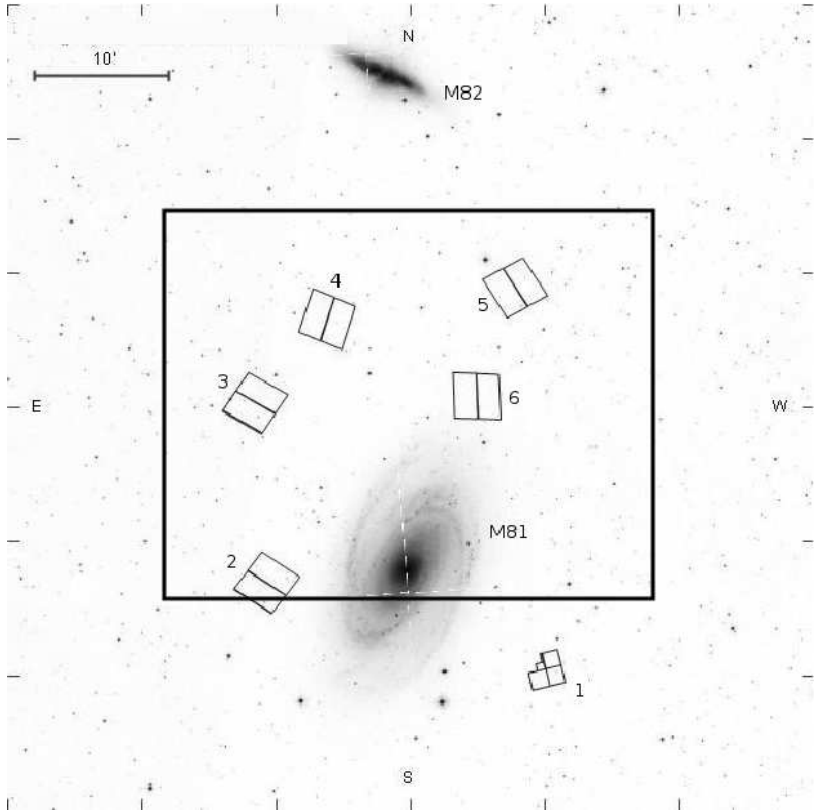}
\caption{One square degree image showing 
the region observed with Suprime-Cam (large rectangle)
relative to M81 and M82.
North is up and east is left.
The legend at upper left gives the
scale of 10 arcmin ($\approx 10$ kpc).
Also shown are the locations of several other 
fields relevant to this paper which were imaged with Advanced Camera for
Surveys or Wide Field and Planetary Camera 2.
These fields are labeled 1 -- 6 and were studied by
(1) \citet{Tikhonov05} and \citet{Mouhcine05}, (2) \citet{Sabbi08} 
and \citet{Weisz08}, 
(3) \citet{deMello08}, (4) and (5) \citet{deJong08}, 
and (6) \citet{Williams08}.
}
\label{fig:fields}
\end{figure*}

The best current method to study the faint outer regions of galaxies
is to resolve their stellar populations.
This allows one to directly probe fainter 
surface brightness levels and to place 
tighter constraints on age and metallicity
than is possible with diffuse light.
For example, \citet{Mouhcine05} studied a sample 
of nearby spiral galaxies with the 
Wide Field and Planetary Camera 2 (WFPC2) on board the 
{\it Hubble Space Telescope} (HST).
They found the red giant branch (RGB) metallicity 
in the outskirts of these galaxies, as inferred from
the colors of their RGB stars, to be positively
correlated with galaxy luminosity, qualitatively
consistent with theoretical expectations for stellar spheroids
in $\Lambda$CDM \citep{Font06} and monolithic
collapse scenarios \citep{Larson74}.
Their conclusions were complicated, however, 
by the small field-of-view of HST 
and the possibility of multiple populations along the
line-of-sight.
Indeed, when selected kinematically, 
halo stars in the three Local Group
spirals do not show a significant correlation
\citep{Chapman06,Kalirai06,Ferguson07,Mouhcine07,Mcconnachie06}.

Ground-based, wide-field mapping of resolved stellar populations
over large portions of nearby galaxies
is an essential complement to pencil-beam HST-based studies.
This approach is necessary for understanding their global
structure, which may exhibit population gradients
and inhomogeneities on a variety of scales.
For example, the INT and MegaCam surveys 
of M31 \citep{Ibata01,Ferguson02,Ibata07} have
revealed a giant stellar stream and multiple other
substructures distributed throughout M31's halo 
which are undetectable with diffuse light and most
easily seen in the spatial distribution of RGB stars.
These substructures 
have sizes and metallicity variations on scales
much larger than HST's field-of-view.

To place the most statistically significant constraints
on theoretical galaxy formation models, we must increase
the number of systems whose individual old and intermediate age
stars have been resolved over large portions of their
outer disks and halos.
This means going beyond the Local Group, a task which has only
recently become possible thanks to the advent of 
wide-field imagers on large telescopes 
\citep[e.g.][]{Bland-Hawthorn05,Davidge08,Davidge08M81}.
At a distance of 3.6 Mpc, 
corresponding to $(m-M)_0 = 27.78$ \citep{Freedman94}, M81 is
the nearest massive spiral like the MW and M31, 
yet it is still close enough to resolve its low-mass 
giant stars. 
With a Hubble morphological class of Sab 
and T-type of 2 \citep{deVauc91}, a total
dynamical mass inside 20 kpc of $\sim 10^{11}\ M_{\sun}$, 
and peak circular rotation velocity $\sim 250\ \rm km\ s^{-1}$
\citep{Gottesman75}, 
M81 is similar in many respects to the MW and M31.
It is therefore an essential system for testing our 
understanding of the formation of massive spirals.

M81 is also the largest member of the nearest interacting
galaxy group.  Its two brightest neighbors, M82 and NGC 3077, 
lie within a projected distance of 60 kpc.
The disturbed nature of the system is most easily visible
when observing the 21 cm emission line of neutral hydrogen.
Maps of the HI content throughout the system show extended tidal 
streams and debris between the galaxies 
\citep{Gottesman75,Yun94}.
The computer simulation of \citet{Yun99} 
explains many of these tidal features 
as being the result of close encounters 
between M81 and each of its neighbors $\sim 200 - 300$ Myr ago.
The stellar contents of the most
prominent debris, Arp's Loop and Holmberg IX (HoIX), 
have been studied with 
WFPC2 by \citet{Makarova02} and with the
Advanced Camera for Surveys (ACS) by \citet{deMello08}, 
\citet{Sabbi08}, and \citet{Weisz08}. 
These authors concluded these objects may be tidal dwarf galaxies, 
new stellar systems that formed in gas stripped from
interacting galaxies.

A few papers have examined the large-scale, young
resolved stellar content around M81 
\citep[e.g.][]{Durrell04,Davidge08M81,Davidge09}.
They identified several groupings of O-B stars and red supergiants
which formed within the last $\sim 100$ Myr and which 
are located in M81's spiral arms and tidal tails.
In this paper, we present wide-field imaging of M81 
obtained with the Suprime-Cam instrument on the 8-m
Subaru telescope.
Our data reach over one magnitude below the RGB tip.
This limiting depth is fainter than previous ground-based
data in M81, and enables a detailed, 
yet global view of {\it both young and old}
stellar structure around this galaxy.
We present evidence for a faint extended structural component
beyond the bright optical disk whose overall
structure and metallicity we can constrain because
we resolve its stellar constituents.
In \S \ref{sec:obs}, we outline the observations and data reduction and,
in \S \ref{sec:cmd}, we present the color-magnitude diagrams of point sources 
and non-M81 contaminants.
We examine the 2-dimensional stellar spatial 
distribution in \S \ref{sec:spatial} 
and the radial surface brightness and stellar density
profiles in \S \ref{sec:radial}.
We discuss possible interpretations in \S \ref{sec:disc} and
summarize the results in \S \ref{sec:summ}.

Throughout this paper, we adopt for M81's disk 
an inclination of $58\deg$, 
and position angle (measured N through E) of $157\deg$.
At a distance of 3.6 Mpc, 1 arcmin $\approx$ 1 kpc in projection.
The $R_{25}$ radius of the galaxy is 13.8 arcmin 
\citep{deVauc91}.

\section{Observations and Data Reduction}
\label{sec:obs}

The observations were obtained with the Suprime-Cam instrument 
\citep{Miyazaki02} on the 8-m Subaru telescope on the nights of January 
7 -- 8, 2005 (S04B, PI=N. Arimoto).  This instrument consists of 10 CCDs of 
2048x4096 pixels arranged in a 2x5 pattern, with a pixel scale of 0.2~arcsec 
and a total field of view of approximately 34x27~arcmin (including long edge 
inter-chip gaps of 16 -- 17 arcsec and short edge gaps of 5 -- 6 arcsec). 

The original aim was to cover M81 using two field centers, one to the north 
and another to the south of its nucleus 
at $(\alpha,\delta) = (9^h 55^m 33^s.2,+69^\circ 03' 55'')$.  
Weather conditions limited the amount 
of good-quality data obtained to the more northerly field centered at 
$(9^h 55^m 30^s.0,+69^\circ 16' 00'')$ (J2000.0).
The large rectangular box in Fig.~\ref{fig:fields} outlines the
size and location of this field relative to M81, to M82, and to several other 
fields imaged with WFPC2 and ACS which are the most relevant to our current 
study of M81's stellar outskirts.  The HST fields are labeled 1 -- 6 and were 
studied by (1) \citet{Tikhonov05} and \citet{Mouhcine05}, (2) \citet{Sabbi08}
and \citet{Weisz08}, (3) \citet{deMello08}, (4) and (5) \citet{deJong08}, 
and (6) \citet{Williams08}.

\begin{figure*}[t]
\epsscale{0.9}
\plotone{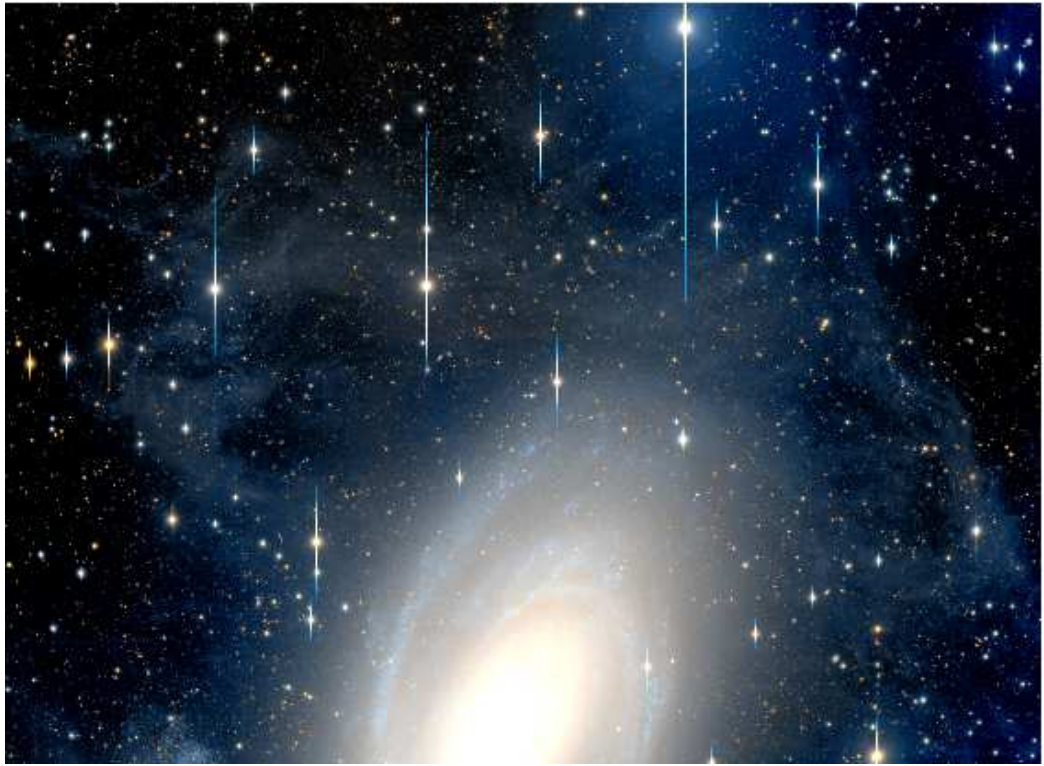}
\caption{Cropped color composite image of M81 created 
from the stacked mosaics.
The intensity scaling is linear and the 
color mapping is similar to that of \citet{Lupton04} with 
$V$ for the blue channel, $i'$ for red, and the average for green.
North is up and east is left.}
\label{fig:mosaic}
\end{figure*}

We obtained a set of 10 images of M81 
in the Johnson $V$ filter with individual 
exposure times of 630s on each night, and 20 images in the Sloan $i'$ filter 
with exposure times of 215s on the night of Jan.\ 7th. All observations were 
recorded under non-photometric conditions through patchy thin cirrus. 
Comparison of expected throughput with subsequent 
calibration based on SDSS photometry for this region indicated that the 
average attenuation for these images was $<$ 10\%. 
The set of 10 $V$-band images obtained on Jan.\ 7th in relatively poor 
seeing ($\sim 1.5\arcsec$) were later rejected from the analysis.
The remaining $V$-band images were taken in an average seeing of 
$\sim 0.7\arcsec$ while the $i'$-band images had an average seeing of
$\sim 1.1\arcsec$.  
To fill in the chip gaps and facilitate the removal of cosmic 
rays and bad pixels, individual images were dithered by $\sim 25$ arcsec,
resulting in a mosaic with total field of view $\approx$ 36x28 arcmin.
Flat-field and interchip gain variations were removed with master flats
obtained by combining 12 and 11 twilight sky flats 
in the $V$ and $i'$ filters, 
respectively. After flat-fielding, remaining large scale variations in
dark sky level, measured directly from stacked dark sky images at several 
different positions obtained during this run, 
were less than 1\% of sky.  
An $i'$-band fringe frame acquired from 
an earlier Suprime-Cam imaging run was used to help assess the degree of 
dark sky fringing present, but this was found to be negligible in our 
data, so this extra image processing step was not required.

After converting the raw data to multi-extension FITS format,
all images and calibration frames were run through a variant of the data 
reduction pipeline developed for the INT Wide Field Survey
\footnote{http://www.ast.cam.ac.uk/\~wfcsur/}.  
The main steps we used in producing a detected 
object catalogue, including background estimation, object detection, 
parameter estimation, and morphological classifcation, are described in more 
detail in \citet{Irwin85,Irwin97}, \citet{Irwin01}, and \citet{Irwin04}.
Here we present a brief overview of these steps.

\begin{figure*}[t]
\epsscale{0.7}
\plotone{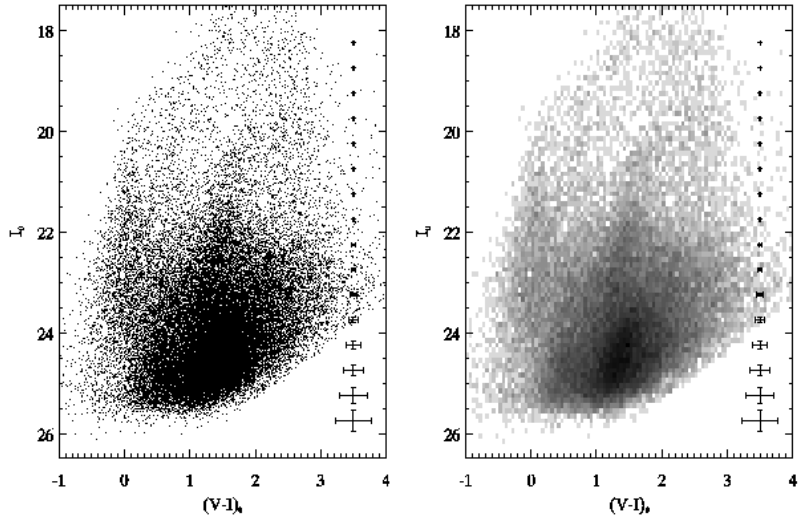}
\caption{Color-magnitude diagram (left) and 
Hess diagram (right) of $\sim 40,000$ point sources
in the Suprime-Cam mosaics of M81.
The Hess diagram is a 2-dimensional histogram 
showing source density on a logarithmic intensity scale.
Error bars on the right-hand side of each panel indicate the
median photometric error at each magnitude.}
\label{fig:cmd}
\end{figure*}

Prior to deep stacking, catalogues were generated 
for each individual processed science image to
both refine the astrometric calibration 
and asses the data quality.
For astrometric calibration, we found that a Zenithal polynomial projection
\citep{Greisen02} provided a good prescription for the World Coordinate System 
(WCS) and included all the significant telescope radial field distortions. We 
used this in conjunction with a 6-parameter linear plate model per detector to 
define the required astrometric transformations.  The 2MASS point source 
catalogue \citep{Cutri03} was used for the astrometric reference system.

The individual image qualities were then assessed using the average 
seeing and ellipticity of stellar images, as well as sky level 
and sky noise 
determined from the object cataloging stage.  As a result of this 
assessment, the set of 
V-band exposures taken on Jan.\ 7th, in much poorer seeing conditions, 
were not included in the final V-band stack.  
Images were
stacked at the detector level using the updated WCS information to accurately
align them with respect to a reference image.  The background level in the
overlap area between each stacked image and the reference was adjusted 
additively to compensate for sky variations during the exposure sequence
and the final stack included seeing weighting, confidence (i.e. variance) map 
weighting, and clipping of cosmic rays.

Next, we generated detector-level catalogues 
from the stacked images and updated
the WCS astrometry in the FITS image extensions prior to mosaicing all 
detectors together.\footnote{
Mosaicing the ensemble to form a single large image minimizes 
the impact of lower total effective exposure times
in the overlap regions between detectors, 
particularly toward their edges.} 
Residual astrometric errors over the whole stacked array 
were typically $< 0.1\arcsec$, greatly simplifying this process.  Slight 
offsets in underlying sky level between the stacked detector images caused 
small (typically $\sim$0.1--0.2\% of sky), but still visible, discontinuities 
in the final mosaics.  These offsets were due to 
small color equation differences in 
the detectors and the relatively blue color of the twilight sky compared to 
dark sky and unresolved diffuse light from M81.  
We corrected these offsets iteratively 
by visual inspection of a 4x4 blocked version of the mosaics using a 
pre-assigned keyword in each relevant detector FITS extension designed for 
this purpose.

Fig.~\ref{fig:mosaic} shows a color composite image constructed from the 
$V$ and $i'$ mosaics, where north is up and east is to the left.  
The image has a linear intensity scaling 
and a color mapping similar to that of \citet{Lupton04} with $V$ for the blue 
channel, $i'$ for red, and the average for green.  The edges were cropped 
due to the slightly different field coverage in the final stacked 
mosaics.

The background light variation over these 
mosaics was rather complex and varied
on relatively short angular scales.  Therefore, 
as an alternative to our default 
method of background estimation and removal,
we also used 
an independent method prior to cataloging in the
crowded central region of M81.  
This step used a filter based on an iterative, non-linear unsharp 
masking and clipping algorithm to progressively 
remove fine details to a user-chosen scale 
length.  An example of the benefits of this approach is shown in Fig.~1 of 
\citet{Mcconnachie07}, where complex background variations due, in
that case, to scattered
light were effectively removed.  Comparison with photometry
generated in a more conventional way demonstrated that even for the brightest
unsaturated stars this aggressive filtering made less than 1\% difference to 
their measured fluxes and had completely negligible effect on fainter stars.
The significant advantages were that if this filtering step was run prior
to mosaicing it obviated the need for detector-level adjustments and it
greatly simplified subsequent object detection and flux estimation.

Object fluxes were estimated using ``soft-edged'' apertures 
of radius $\approx$ FWHM,
which, for relatively uncrowded images, 
deliver 80-90\% of the signal-to-noise of a perfect
PSF fit in a fraction of the time
\citep[e.g.,][]{Irwin97,Naylor98}.  A significant portion of the
detected objects in these images were background galaxies, for which aperture 
photometry was more useful than PSF-fitting.  
Therefore, a series of apertures ranging from 1/2 to 
4 times the FWHM were additionally used as the basis for the morphological 
classification in a curve-of-growth analysis.  We also used these 
apertures to 
compute stellar aperture corrections for the primary flux measure. 
The PSF varies in a rather subtle way over such wide-field images, 
and even though aperture photometry is not as sensitive to this
variation as PSF-fitting, 
we still used the different aperture fluxes to check and correct for it.

Beyond M81's bright optical disk,
the typical object separation increases from $\sim 5\arcsec$
at a deprojected radius $R_{dp} = 14$ arcmin to 
$\sim 10\arcsec$ at $40$ arcmin.
To mitigate the effects of crowding, 
which are more important at $R_{dp}~<~14$~arcmin, 
the pipeline employed two extra refinements to the flux measurement.  
First, the object detection information for neighboring 
regions was used to flag those pixels corrupted by external groups of 
detected objects.  Second, object fluxes from a pixel 
image blend of overlapping 
sources were solved for simultaneously via least-squares fitting of the
aperture fluxes, i.e.\ equivalent to a top-hat shaped PSF, using fixed input 
coordinates.  With this approach, bad pixels and regions of extremely low 
confidence were also readily flagged and avoided appropriately.  Direct 
comparison of the pipeline-generated photometry 
with full PSF-fitting over a broad range of 
images for other projects has shown that in all 
but the most densely crowded regions, like
the centers of globular clusters, similar quality results are achieved.

Due to the
non-photometric observing conditions, we brought our flux measurements 
onto a Vega-like system (Johnson $V$ and Cousins $I$) 
by matching stars in our catalog with those in the 
SDSS, after applying the Lupton transformation 
equations provided on the SDSS website 
\footnote{http://www.sdss.org/dr6/algorithms/sdssUBVRITransform.html}.
Only stars with photometric errors in the transformed system $< 0.2$ mag
were used in the matching.  The resulting photometric zero-point is accurate 
to $\sim 0.05$ mag, which is adequate for our study.

The final catalog 
contains 
$\sim$ 40,000 objects with stellar, or probable stellar, classifications in 
both passbands (hereafter referred to as point sources), meaning their image 
shape parameters lie within $2-3\sigma$ of the main stellar locus.  Relaxing 
this criterion in $I$ makes little difference to our results, but relaxing it 
in $V$ actually increases contamination from unresolved background galaxies 
since the seeing in $V$ was significantly better.  
The morphological classification algorithm categorized $\sim 23,000$
objects as background galaxies, which we discuss in more detail below.

\section{Color-Magnitude Diagram}
\label{sec:cmd}

Fig.~\ref{fig:cmd} shows
the final point source catalog as a color-magnitude diagram (CMD)
in the left panel and as a Hess diagram in the right panel.
The Hess diagram is a 2-dimensional histogram showing
the source density on a logarithmic
intensity scaling.
There are several clear stellar sequences visible, which 
we discuss in more detail below.
On the right-hand side of each panel are the 
median photometric errors 
reported by the data reduction pipeline for all point sources 
in 0.5 mag-wide bins, where the color errors are the
V- and $I$-band errors added in quadrature.  
The median $I$-band error is 0.1 at $I \approx 24.7$
and $\approx 0.14$ at $I$ = 25 while the median color error
is 0.1 at $I \approx 24.3$ and $\approx 0.16$ at $I$ = 25.0.  
These errors reflect Poisson noise in the sources 
and the sky level.  
They do not account for systematic color and magnitude shifts 
or for correlated errors in both bands, which 
can occur in heavily crowded regions, 
but by restricting most of the point source analysis to 
M81's outskirts, we do not expect this to be a serious issue.
We also note that point sources are saturated at $V \lesssim 19$ 
and $I \lesssim 19$.

The extinction maps of \citet{Schlegel98} indicate 
a variable extinction across our field, with $\sim 15\%$
of point sources having $E(B-V)$ values at least 0.01 mag in excess of the
median (0.08) and a total range of 0.05 -- 0.12.
Henceforth, we apply extinction corrections on a
star-by-star basis using the \citet{Schlegel98} maps and
\citet{Cardelli89} extinction law, for which $R_V = 3.1$
and $A_I/A_V = 0.479$.
\citet{Schlegel98} removed extragalactic objects from their maps 
down to a certain flux limit and replaced them with median 
values from the surrounding sky.
By examining their mask maps, one can see that the inner
$\sim 16$ arcmin in projected radius around M81 
(including HoIX and part of Arp's Loop) was indeed
processed in this manner.
Thus, the star-by-star correction does not 
include extinction internal to M81, 
but this should not be a serious problem since we focus
on the RGB stars, which should be the
least affected by internal extinction \citep{Zaritsky99}, and our CMD 
selection boxes (described below) are larger than
the expected amount of internal extinction.
For instance, the spectro-photometric study of M81's
diffuse light by \citet{Kong00} found an average 
$E(B-V) \sim 0.2$ throughout M81's bright optical disk, 
suggesting there may be $\sim 0.12$ mag of 
internal reddening within this region.

\begin{figure}[t]
\epsscale{0.7}
\plotone{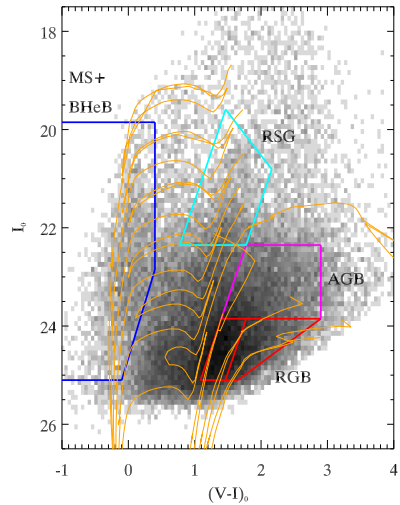}
\caption{Hess diagram of $\sim 40,000$ point sources with theoretical
isochrones from \citet{Marigo08} overlaid.
The young isochrones at $(V-I) \sim 0$ 
have ages of 10.0, 17.8, 31.6, 56.2, 
100, and 178 Myr and a metallicity [M/H] $= -0.4$.
The three old isochrones at $(V-I) \sim 1 - 3$ have a common
age of 10 Gyr and [M/H] = --1.3, --0.7, and --0.4.
The boxes are used to select M81 stars in different
evolutionary stages: main sequence and blue helium burning (MS+BHeB),
red supergiant (RSG), asymptotic giant branch (AGB), and red giant
branch (RGB).
The 50\% completeness level is estimated to lie at $I_0 \sim 25.0$ for
MS+BHeB stars and $I_0 \sim 24.4$ for RGB stars 
(see \S \ref{sec:cmd} for details).}
\label{fig:cmdiso}
\end{figure}

In Fig.~\ref{fig:cmdiso}, we show the point source 
Hess diagram 
overlaid with isochrones from \citet{Marigo08} with the 
circumstellar dust option turned off.  
The young isochrones at $(V-I) \sim 0$ 
have ages of 10.0, 17.8, 31.6, 56.2, 
100, and 178 Myr and a metallicity 
[M/H]~$= {\rm log}(Z/Z_{\sun}) = -0.4$.
We use this metallicity as a fiducial value
because the young stars sampled by our field
exist in a variety of environments with 
[M/H] ranging from $\sim -0.7$ to 0.0
\citep{Zaritsky94,deMello08,Sabbi08,Davidge08M81}.
The three old isochrones at $(V-I) \sim 1 - 3$ have a common
age of 10 Gyr and [M/H] = --1.3, --0.7, --0.4.
The discontinuities in the asymptotic giant branch (AGB) are explained 
in \citet{Marigo08} and are caused by
changes in the opacity tables at the
transition to the thermally pulsing phase.

In what follows, we will focus on several particular CMD regions
(outlined in Fig.~\ref{fig:cmdiso}) which isolate stars in different
evolutionary stages at M81's distance.
The blue lines in Fig.~\ref{fig:cmdiso} mark the region occupied by 
main sequence stars $\lesssim 20$ Myr old 
and by blue helium burning stars at the hottest extension of
the blue loop phase (MS+BHeB) with ages $\lesssim 100$ Myr.
Stars within the cyan polygon are red supergiants (RSGs)
with ages in the range $\sim 20 - 200$ Myr.
The tip of the RGB lies at $I \sim 24$, 
so stars within the red lines are mainly RGB stars
with ages $\sim 1 - 10$ Gyr.
Note that there could be some contamination 
of the RGB box by young, red 
helium burning stars with masses of $\sim 3-4\ M_{\sun}$, 
particularly if they have [M/H] $> -0.4$.
The magenta lines enclose AGB stars
above the RGB tip, 
which tend to have somewhat younger ages ($\sim 0.5 - 8$ Gyr)
than the RGB stars
\citep{MartinezDelgado99,Gallart05}.

We have also divided the RGB box into 
``metal-poor'' ([M/H] $\lesssim -0.7$) and 
``metal-rich'' ([M/H] $\gtrsim -0.7$) subregions.
There could 
be some overlap in the metallicities probed by these
subregions, but they are broader in color than the photometric
errors, and so are useful in identifying any population gradients.
Some of the most metal-poor ([M/H] $\lesssim -1.3$) and
metal-rich ([M/H] $\gtrsim -0.4$) RGB stars
may fall outside the total RGB box, but
extending it further to the blue or red would
increase contamination from MW foreground stars and 
unresolved background galaxies, 
and increase uncertainties due to incompleteness.
Comparing the number of RGB stars with [M/H] $\gtrsim -0.4$
in the \citet{Williams08} CMD 
(field 6 in Fig.~\ref{fig:fields} at $R_{dp} = 14$ arcmin)
to the number in
our Subaru CMD reveals that most of these stars are too faint
to be detected in our data.
However, our detailed RGB analysis focuses on exterior regions where
significantly fewer metal-rich RGB stars are expected to exist, 
a point substantiated in \S \ref{sec:radial}.

\begin{figure}[t]
\epsscale{0.7}
\plotone{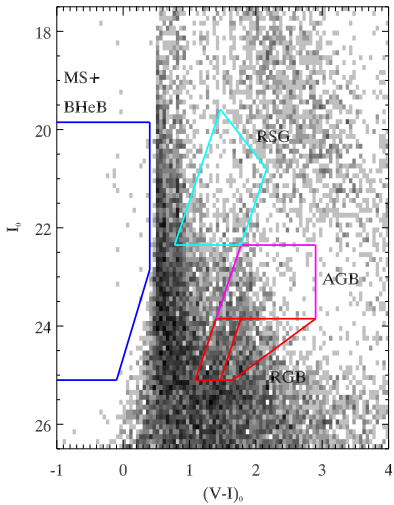}
\caption{Hess diagram of foreground stars
based on the Besan\c{c}on model of the Milky Way \citep{Robin03}
for a field with the same area and line of sight as the M81 mosaics.
The stars were scattered according
to the photometric errors of the real data, but no
completeness corrections were applied.
The number of predicted foreground stars is $\sim 2\%$ of the 
number of point sources over the observed magnitude range.
The selection
boxes (from Fig.~\ref{fig:cmdiso}) avoid the most heavily
contaminated regions.}
\label{fig:cmdfground}
\end{figure}

\begin{figure}[t]
\epsscale{0.7}
\plotone{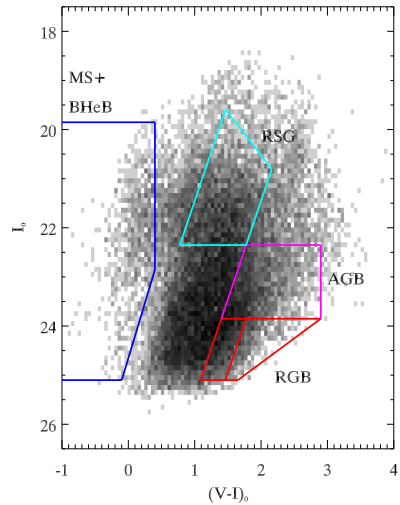}
\caption{Hess diagram 
of $\sim 23,000$ extended objects in the M81 field.
The selection
boxes (from Fig.~\ref{fig:cmdiso}) avoid the most heavily
contaminated region.  The surface density
of extended objects in the total red giant branch box is less than 
half that of the point sources in Fig.~\ref{fig:cmdiso}
at all radii.}
\label{fig:extended}
\end{figure}

\begin{figure}[t]
\epsscale{0.7}
\plotone{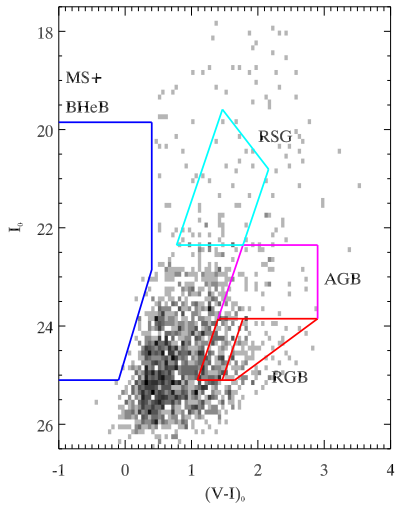}
\caption{Hess diagram of $\sim 2,000$ contaminants 
extracted from an area 112 arcmin$^2$ 
($\sim 10\%$ the total field-of-view around M81) 
at vertical heights $\gtrsim 20$ kpc from the plane of NGC 4244.
These objects are mostly unresolved background galaxies.
The selection
boxes (from Fig.~\ref{fig:cmdiso}) avoid the most heavily
contaminated region.
Table \ref{tab:bg} lists the number counts of objects 
in the boxes, which are used to subtract contaminants 
from the radial stellar
density profiles of M81.}
\label{fig:cmdbground}
\end{figure}

As can be seen in Fig.~\ref{fig:cmdfground} -- 
\ref{fig:cmdbground}, these selection
boxes sample regions of the CMD that are likely
to maximize the number of M81 stars relative to 
the number of foreground MW stars and background galaxies.
Fig.~\ref{fig:cmdfground} shows the foreground
star Hess diagram predicted by 
the Besan\c{c}on model \citep{Robin03} for a field with 
the same area and line of sight as ours.  
The stellar colors and magnitudes have
been scattered using a simple exponential function to mimic
the increase of photometric error with magnitude seen in the real data.  
The Besan\c{c}on model predicts that the number of foreground stars
is $\sim 2\%$ of all point sources
over the magnitude limits of the observed CMD. 
Since applying completeness corrections would only 
decrease this already small percentage, 
we do not use this model to correct 
the stellar number counts for MW stars.

At bright magnitudes ($I \lesssim 23$), morphological 
classification effectively removes background galaxies, 
but they can still be a significant source
of contamination at faint magnitudes, especially 
when observing from the ground.
To check the effectiveness of our classification algorithm, 
Fig.~\ref{fig:extended} shows the Hess diagram 
of the $\sim 23,000$ objects classified as extended.
The CMD selection boxes are overlaid to facilitate comparison
with the other CMDs presented here.
The extended objects are concentrated in 
a broad diagonal band, the bulk of which lies at bluer
colors than the RGB and AGB selection boxes.
Importantly, the extended objects are distributed
differently from the point sources, 
with no obvious stellar sequences aside from some
misclassified MS+BHeB stars located in spatially crowded 
regions where their PSFs overlap significantly.
This fact suggests that the algorithm
has correctly classified the majority of detected M81 stars.
The surface density of point sources within the total RGB box
is over 2 times higher at all radii than the surface density of
extended objects in the same box.
Nevertheless, some compact background galaxies can be
mistaken for point sources and these contaminants 
become increasingly important at large galactocentric distances.

The large spatial extent of M81 and various tidal features
throughout Suprime-Cam's field-of-view make it difficult to 
estimate the contamination from background 
galaxies misclassified as point sources using
just the M81 mosaics themselves.
However, we can estimate this level from
images of the edge-on spiral galaxy, NGC 4244, 
taken during the same observing run.
Fig.~\ref{fig:cmdbground} shows the 
Hess diagram of a control field extracted from 
an area of 112 arcmin$^2$ at 
vertical heights $\gtrsim 20$ kpc above NGC 4244's disk, 
assuming a distance of 4.4 Mpc \citep{Seth05a}.
This area corresponds to $\sim 10\%$ the 
total field-of-view around M81. 
Extinction corrections were applied to each object individually 
using the \citet{Schlegel98} maps, which indicate
a median $E(B-V)$ of 0.02.
The control field has $\sim 2,000$ sources distributed 
in the CMD in a similar manner as the extended objects 
of the M81 field.
The most densely populated region lies in the same diagonal band 
at $0 \lesssim (V-I)_0 \lesssim 1$.
The CMD selection boxes avoid this region 
while still sampling as much of M81's stellar populations
as possible.
The surface densities of objects within the boxes 
in the contaminant CMD 
are listed in Table \ref{tab:bg} with Poisson errors.
The contaminant surface density in the total RGB box
is less than the corresponding point source density at all radii, 
reaching a maximum fraction of $\sim 50\%$ 
at a deprojected radius $R_{dp} \sim 37$ arcmin.

We also estimated background contamination by
downloading from the HST archive two M81 fields
observed with ACS 
(SNAP 10523, PI: de Jong) at $R_{dp} \sim 20 - 30$ arcmin.
These two fields were observed in the $F606W$ and $F814W$
filters and are indicated 
as fields 4 and 5 in Fig.~\ref{fig:fields}.
We performed the data reduction using the ACS module of the DOLPHOT package 
\footnote{DOLPHOT is an adaptation of the photometry 
package HSTphot \citep{Dolphin00}. It can be downloaded 
from http://purcell.as.arizona.edu/dolphot/.}
following the basic steps outlined in the DOLPHOT manual.
We defined objects as point sources if they were 
classified by DOLPHOT as 'good stars' with 
$S/N > 5$ and crowding parameter $< 0.5$
in both filters, and if the overall $\vert sharp\vert < 0.1$
and $\chi < 3$.
Magnitudes in the ACS filter system were transformed to
the Johnson-Cousins system using the equations in
\citet{Sirianni05}.
We matched point sources in our Subaru catalog with point sources 
in the ACS catalogs
by applying small, constant offsets in right ascension and declination
and then counted the number of unmatched point sources in
our catalog, presumed to be unresolved background galaxies.  
Note that this method assumes the ACS point-source catalogs themselves have
no contamination from unresolved background galaxies.
Reassuringly, the resulting background surface densities in each of the
CMD selection boxes (see Table \ref{tab:bg}) were 
consistent with those derived from the NGC 4244 data to within $\sim 2\sigma$.
Compared to the NGC 4244 background, the ACS background 
is $\sim 50\%$ higher in the total RGB box, but the ratio of 
metal-poor/metal-rich RGB background is the same.
The ACS background is also higher in the MS+BHeB box, but
lower in the AGB and RSG boxes.

\begin{figure}[t]
\epsscale{0.95}
\plotone{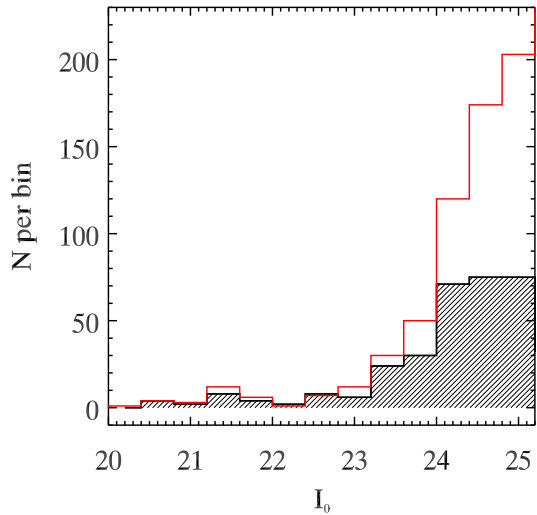}
\caption{The dereddened luminosity function
of point sources in two combined ACS fields 
(open histogram; fields 4 and 5 in Fig.~\ref{fig:fields})
compared to that of the matched point sources
in our catalog (hatched histogram).
The ratio of the two luminosity functions
indicates that our catalog is $\gtrsim 50\%$ complete
for $I_0 < 24.4$ at deprojected radii $R_{dp} = 20 - 30$ arcmin.}
\label{fig:lumfuncs}
\end{figure}

\begin{deluxetable}{lcc}[b]
\renewcommand{\arraystretch}{1.0}
\tablecaption{Surface densities of contaminants.}
\tablewidth{0pt}
\tablehead{\colhead{CMD} &\colhead{NGC4244} &\colhead{ACS} \\
\colhead{region} &\colhead{($\rm arcmin^{-2}$)} &\colhead{($\rm arcmin^{-2}$)}}
\startdata
 MS+BHeB        & $0.03  \pm 0.02 $ & $0.3  \pm 0.2 $ \\
 RSG            & $0.37  \pm 0.06 $ & $0.3  \pm 0.2 $ \\
 AGB            & $0.52  \pm 0.07 $ & $0.4  \pm 0.2 $ \\
 Metal-poor RGB & $1.28  \pm 0.11 $ & $2.0  \pm 0.4 $ \\
 Metal-rich RGB & $0.64  \pm 0.08 $ & $1.0  \pm 0.3 $ \\
 Total RGB            & $1.93  \pm 0.13 $ & $3.0  \pm 0.5 $ \\
\enddata
\label{tab:bg}
\end{deluxetable}

We decided to use the NGC 4244 data to subtract the background from
the radial density profiles presented below, 
since they sample an area 5 times greater than
the ACS fields, are free from the uncertainties in the
object matching, and rely 
on the same instrument, filters, 
data reduction procedure, and morphological classification 
as the M81 data presented here.
This approach assumes a negligible contribution from stars
and globular clusters in 
NGC 4244's halo 20 kpc above the plane 
and that the background 
galaxy and foreground star populations (at the magnitudes we are probing)
are statistically equivalent to those around M81.
At $I \sim 23 - 25$, the
background galaxies have a median $z \sim 1$ \citep{Ilbert06}.
At this redshift, $1 \arcsec \approx 8$ kpc, and the entire Suprime-Cam
field-of-view samples a projected volume $\sim 15$ Mpc wide, 
large enough to gather a statistically
representative sample of background galaxies
regardless of position on the sky.
Cosmic variance may be 
more of a concern here, since we are not able
to use Suprime-Cam's entire field-of-view.
However, cosmic variance is more likely to affect the ACS
background, which covers much less area.
With regards to foreground contamination, 
NGC 4244 is separated from M81 by $\approx 40$ deg
and lies at a higher Galactic latitude, 
but both galaxies are 
significantly above the MW plane. 
Indeed, the Besan\c{c}on model predicts only $\sim 10 - 25\%$ fewer
foreground stars towards NGC 4244 than towards M81, 
within the CMD selection boxes.
Our general conclusions are unchanged if we adopt the
ACS background estimates, but to be conservative, 
we adopt the differences between the ACS and NGC 4244
estimates as systematic errors, 
which are added in quadrature with the Poisson noise in the latter.

Another important issue is estimating the 
completeness rate of our Subaru catalog.
The standard method is
to run artificial star tests, in which thousands of stars
with known magnitudes and colors are inserted into the
original images, which are then re-processed.
However, this method requires detailed knowledge of the PSF
and how it varies across the field-of-view.
Since we have used optimized
aperture photometry that is insensitive to these variations, 
and given the complexities of the mosaicing and
data reduction, we have elected not to run 
artificial star tests.
Instead, we have used two alternative methods to
probe the completeness rate of our catalog.

First, the completeness rate can be estimated 
by comparing the number of point sources in 
the Subaru and ACS catalogs.
Fig.~\ref{fig:lumfuncs} shows the dereddened luminosity functions
of all point sources in the two ACS background fields (open histogram)
and the matched point sources in the Subaru catalog (hatched histogram).
This figure shows that in the range $R_{dp} \sim 20 - 30$ arcmin 
and for $I_0 < 24.4$, 
the Subaru catalog is $\gtrsim 50\%$ complete 
relative to the ACS catalog.
In addition, the number counts in the CMD selection boxes
indicate that each box is $\gtrsim 50\%$ complete
relative to the ACS catalog, 
with a statistical uncertainty of 5\% or more
depending on the number of sources in each box.

Second, we use the Subaru luminosity function (LF) peak
as a proxy for the $50\%$ completeness level.
It is only an approximate indicator of this level
because it depends on the true shape of the LF.
Nevertheless, for our entire catalog, this peak 
occurs at $I_0 \sim 24.45$, 
consistent with the above estimate.
Table \ref{tab:lfpeak} lists the LF peak magnitude 
for several different color ranges.
This table shows that all of the MS+BHeB and
RSG boxes and most of the AGB box lie brighter than 
the LF peak, while about
half the metal-poor RGB box and most of the
metal-rich RGB box lie fainter than the peak.
We note that restricting our analysis to stars
brighter than the LF peak makes no significant
difference to our conclusions, but we use
the entire CMD boxes, anyway, to decrease Poisson noise.
Thus, one should be mindful of the different completeness
levels for the metal-poor and metal-rich RGB stars
when examining their relative distributions.

\begin{deluxetable}{lcc}[b]
\renewcommand{\arraystretch}{1.0}
\tablecaption{Luminosity function peak magnitudes.}
\tablewidth{0pt}
\tablehead{\colhead{$(V-I)_0$} &\colhead{LF peak} \\
\colhead{range} &\colhead{$I_0$ magnitude} }
\startdata
-1.0 - 0.5   & 25.15 \\
 0.5 - 1.0   & 24.85 \\
 1.0 - 1.5   & 24.55 \\
 1.5 - 2.0   & 24.45 \\
 2.0 - 2.5   & 24.05 \\
 2.5 - 3.0   & 23.35 \\
\enddata
\label{tab:lfpeak}
\end{deluxetable}

\section{Two-Dimensional Spatial Distribution}
\label{sec:spatial}

\begin{figure}[t]
\epsscale{0.95}
\plotone{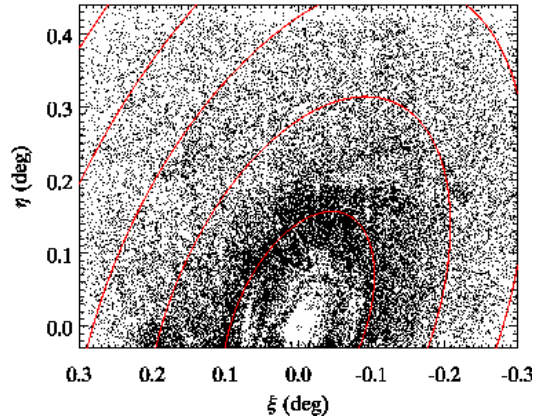}
\caption{Tangent plane projection of point sources with $I_0 < 25.15$
and $V_0 < 26.85$.  Ellipses denote deprojected
radii of 10, 20, 30, 40, and 50 arcmin (1 arcmin $\approx 1$ kpc).
No correction for contaminants has been made.
The diffraction spikes of a few highly saturated stars appear as thin 
vertical white stripes.
}
\label{fig:mapall}
\end{figure}

Fig.~\ref{fig:mapall} displays the 
tangent plane projection 
of all point sources
with $I_0 < 25.15$ and $V_0 < 26.85$.
The ellipses correspond to deprojected 
radii of $R_{dp} =$ 10, 20, 30, 40, and 50 arcmin 
(recall that $\rm 1\ arcmin \approx 1\ kpc$).
No correction for contaminants has been made.
The diffraction spikes of a few highly saturated stars appear as thin 
vertical white stripes.
Visual inspection reveals significant substructure 
associated with spiral arms in M81's disk and
the tidal features, HoIX at $(\xi,\eta) \sim (0.2,-0.02)$
and Arp's Loop at $\sim (0.15,0.25)$.
The hole in the central region is due to saturation and stellar crowding.

\begin{figure*}[t]
\epsscale{0.95}
\plotone{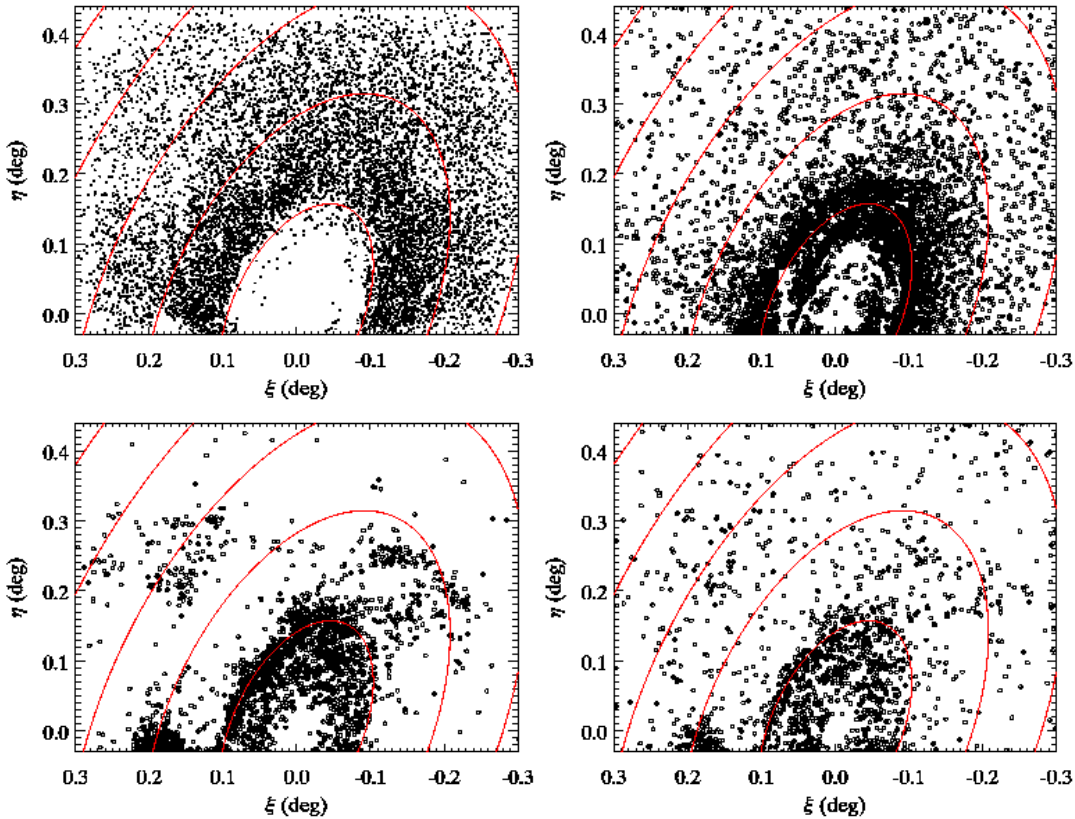}
\caption{Going clockwise from top left, the 
tangent plane projection of red giant branch stars (ages $\sim 1 - 10$ Gyr), 
asymptotic giant branch stars (ages $\sim 0.5 - 8$ Gyr), 
red supergiant stars (ages $\sim 20 - 200$ Myr), and
main squence and blue helium burning stars (ages $\lesssim 100$ Myr).
Ellipses denote deprojected
radii of 10, 20, 30, 40, and 50 arcmin (1 arcmin $\approx 1$ kpc).
No correction for contaminants has been made.}
\label{fig:maprgb}
\end{figure*}

Going clockwise from top left, Fig.~\ref{fig:maprgb} shows the spatial
maps of RGB, AGB, RSG, and MS+BHeB stars.
No correction for contaminants has been made.
From these maps, one can see that most of the overdensities
observed in Fig.~\ref{fig:mapall} are due to star formation
in the last 100 Myr.
There are no stellar groupings in the RGB and AGB maps that can be
unequivocally linked to groupings in the MS+BHeB or RSG maps.
Indeed, the RGB and AGB stellar distributions are smooth
except near the chip edges and mosaic corners, 
where the completeness rate is lower due to 
different field centers and vignetting.
These regions are therefore excluded from the radial
stellar density profiles discussed in \S\ref{sec:radial}.

In their ACS images of Arp's Loop at (0.16,0.23), 
\citet{deMello08} found that the RGB stars
were broadly distributed like the youngest stars,
except for a couple O-B associations, 
which may be too young to have formed RGB stars.
Our stellar maps, which reveal the stellar distribution
on a larger scale than the HST images, appear to contradict this finding
because there is no clear broad overdensity of RGB stars
associated with Arp's Loop.
This fact is consistent with the idea that 
Arp's Loop was initially entirely gaseous tidal debris that began forming
stars $\sim 200 - 300$ Myr ago, around the time 
of the last close encounters between
M81, M82, and NGC 3077.
It is possible that young, red He-burning stars were 
contaminating the \citet{deMello08} RGB sample, 
but there is no obvious reason why their
sample would suffer more from this contamination than ours.
It is also possible that the completeness rate
in our maps is not exactly uniform across this part of Arp's Loop.
This is obviously the case for HoIX, where the crowding is so severe
that it appears as a hole in the RGB star map, so we
are unable to comment on the assertion of \citet{Sabbi08}
that most of its RGB stars belong to M81.

One of the stellar concentrations identified by \citet{Davidge08M81}, 
TDO 3, lies within our field-of-view at (-0.13,0.25)
and appears as an overdensity of MS+BHeB and RSG stars.
TDO 3 is clearly associated with the northernmost
spiral arm, which may be tidally driven \citep{Yun94}, but
neither of these objects necessarily are, or will become, distinct tidal 
dwarf galaxies, as may
be the case for Arp's Loop and HoIX.
The CMD of TDO 3 suggests an age spread
of several tens of Myr with the youngest stars $\sim 20$ Myr old, 
consistent with the estimates of \citet{Davidge08M81}.
We are also able to confirm his tentative detection 
of a stellar grouping east of Arp's Loop 
at (0.25,0.24), with a similar age as TDO~ 3.

\begin{figure}[t]
\epsscale{0.95}
\plotone{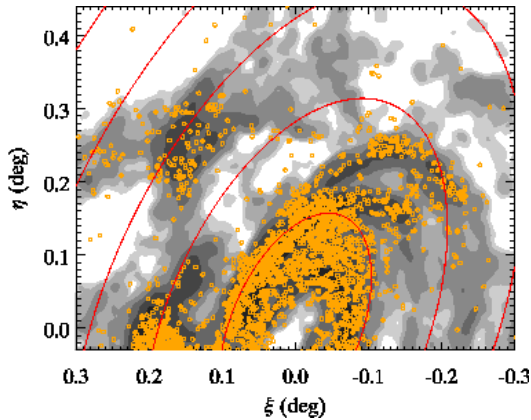}
\caption{Tangent plane projection of main sequence and
blue helium burning stars 
(ages $\lesssim 100$ Myr).  Ellipses denote deprojected
radii of 10, 20, 30, 40, and 50 arcmin (1 arcmin $\approx 1$ kpc).
No correction for contaminants has been made.
Contours show the column density of HI 
in the VLA mosaic of \citet{Yun94}.
The contour levels 
increase by factors of two with the lowest (white)
covering $< 2 \times 10^{20}\ \rm cm^{-2}$ 
(or $1.7~M_{\sun}~\rm pc^{-2}$).
}
\label{fig:mapyoung}
\end{figure}

Fig.~\ref{fig:mapyoung} shows the distribution of MS+BHeB
stars overlaid on the HI column density map of \citet{Yun94}.
The contour levels increase by factors of two with the
lowest (white) covering densities $< 2 \times 10^{20}\ \rm cm^{-2}$
(or $1.7~M_{\sun}~\rm pc^{-2}$).
The young stars closely trace the HI for column densities
$\gtrsim 8 \times 10^{20}\ \rm cm^{-2}$.
Few young stars exist in regions with lower column density.
This seems consistent with a simple star formation 
threshold gas surface density, but
it could also be caused by a star formation rate in these
low-density regions that is 
too small to produce the high-mass stars we
have resolved.

\begin{figure}[t]
\epsscale{0.95}
\plotone{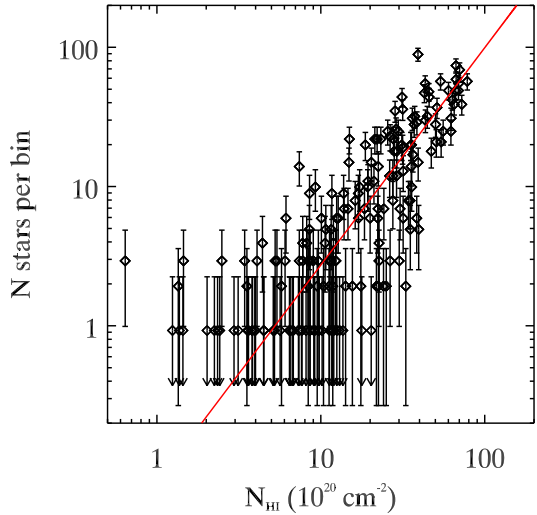}
\caption{Background-subtracted surface density of 
main sequence and blue helium burning stars as a function
of mean HI column density in square bins of width 1.7 arcmin 
($\approx 1.7$ kpc).
Error bars include Poisson noise and systematic error
in the background.
Downward pointing arrows indicate those regions with only one young star.
The line corresponds to a best fit slope of $1.6 \pm 0.2\ (2\sigma)$.
Increasing or decreasing the area of each bin by 
a factor of 4 increases or decreases the slope by 
$\sim 0.1 - 0.2$, respectively.
The measured slope is similar to 
the canonical disk-averaged Kennicutt-Schmidt relation, which 
has a slope of 1.4 \citep{Kennicutt98}.}
\label{fig:ks}
\end{figure}

To investigate the correlation between HI and MS+BHeB stars 
in more detail, 
we divided the field into square bins 1.7 arcmin
($\approx 1.7$ kpc) on a side 
and counted the number of these stars in each bin.
For reference, the beam size of the radio observations
was $\sim 1$ arcmin.
Fig.~\ref{fig:ks} shows the background corrected 
number counts as a function of mean HI
column density within each region.
Errors with downward pointing arrows indicate regions with only
one young star.
A maximum likelihood fit yields a best fit slope of 1.6 
and 2$\sigma$ confidence interval of $\pm 0.2$.
Increasing or decreasing the area of each bin by 
a factor of 4 increases or decreases the slope by 
$\sim 0.1 - 0.2$, respectively.

Our measured slope is roughly consistent with expectations from 
the canonical Kennicutt-Schmidt (KS) relation, which 
has a slope of 1.4 \citep{Kennicutt98}, 
but that relation was derived for disk-averaged
quantities rather than individual regions.
Our slope is within the range (1.2 -- 3.5) 
derived for small regions or azimuthally averaged rings in 
disk galaxies \citep{Wong02,Boissier03,Kennicutt07}.
Recent theoretical work on star formation in 
galactic disks predicts the KS relation to steepen
at low gas densities, due to a star formation
threshold, decreasing molecular fraction, or
flaring of the gas disk \citep{Schaye08,Robertson08}.
Since we have neglected molecular gas, a 
detailed comparison to these studies is unwarranted.
We simply note that our data are consistent with a single
slope down to gas densities of $\sim 2 \times 10^{20}\ \rm cm^{-2}$, 
lower than typically probed using $H\alpha$ 
flux as a star formation tracer and lower than most estimates
of the star formation threshold, which include molecular gas.
Note also that incompleteness may cause some of the fainter MS+BHeB stars
to be preferentially missed in the most crowded regions
with the highest gas densities, possibly biasing our
measured slope downward.
Nevertheless, it is encouraging that we recover a 
reasonable slope using an independent method sensitive to a 
wider age range than $H\alpha$ flux (100 Myr vs.\ 10 Myr).
\section{Radial Profiles}
\label{sec:radial}

\begin{figure}
\epsscale{0.95}
\plotone{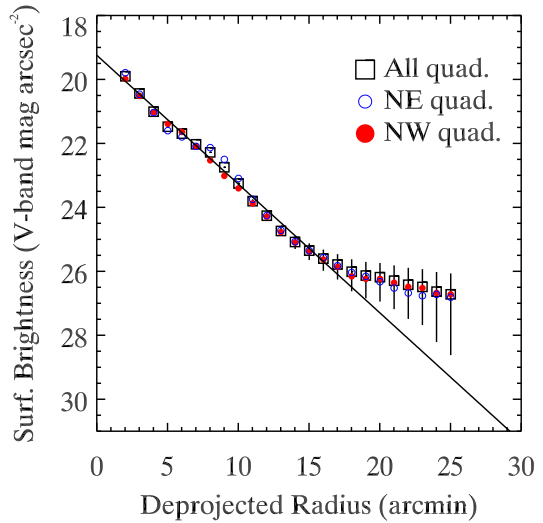}
\caption{V-band surface brightness profile of M81 
derived from diffuse light (squares).
Error bars are $1\sigma$ and include read noise
and uncertainties in the total signal and sky.
Filled and open circles represent, respectively, 
the NW and NE quadrants 
following the convention where north is up and east is to the left
of M81's nucleus.
The solid line is an exponential fit
with a scale-length $h = 2.7 \pm 0.1$ arcmin.
The profile shows evidence for a flattening 
at $R_{dp} \sim 17$ arcmin.}
\label{fig:sbprofile}
\end{figure}

We derived the V-band diffuse light surface brightness profile of M81
using the IRAF ellipse task with elliptical 
annuli of constant center, position angle, and inclination
and after masking saturated stars.
In each elliptical annulus, the median pixel
value was computed after 
two $5\sigma$ clipping iterations.
The sky value was estimated as the average median pixel value of
13 3x3 arcmin boxes near the edges of the mosaic.
Squares in 
Fig.~\ref{fig:sbprofile} 
show the sky-subtracted profile.
The error bars include read noise, Poisson noise
in the total signal, and the standard deviation of the
13 median sky values.
Open circles represent the NE quadrant 
and filled circles represent the NW quadrant
(relative to M81's nucleus and following the 
convention where north is up and east is to the left).
The error bars on the circles are omitted for clarity.
The median extinction of all point sources, 
$A_V = 0.25$, was subtracted from the profiles.
Inside $\sim 2$ arcmin, the bulge is saturated, so
there are no data points there.

The diffuse light surface brightness profile traces M81's 
exponential disk out to $R_{dp} \sim 17$ arcmin, 
farther than previous studies \citep{Kent87,Mollenhoff04,Willner04}.
The exponential scale-length inside 17 arcmin 
(shown by the solid line) is $2.7 \pm 0.1$ arcmin, 
in good agreement with 2.53 arcmin found by Mollenhoff (2004), 
who performed a bulge-disk decomposition 
inside $R_{dp} \sim 9$ arcmin and over all
4 quadrants.
The bump in the NE profile at $\sim 9$ arcmin is due to a
spiral arm.  
There is a clear flattening in the profiles
at $\sim 17$ arcmin, but the $1\sigma$ uncertainties
make the precise location and amount of flattening somewhat ambiguous. 

\begin{figure}
\epsscale{0.95}
\plotone{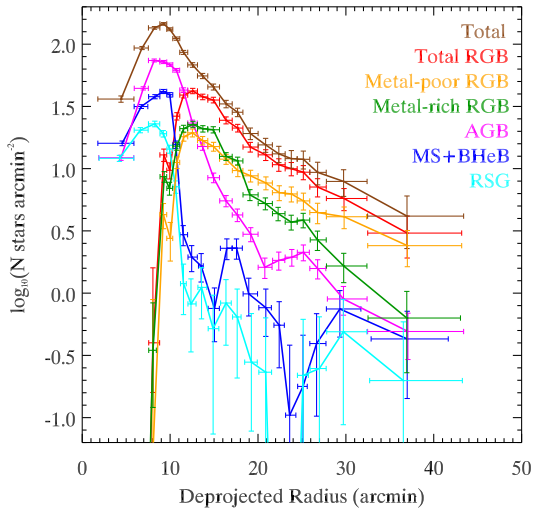}
\caption{Background-subtracted radial density 
profiles for stars in the
different color-magnitude selection boxes.
Vertical error bars include Poisson noise
and systematic uncertainty in the background.
Horizontal error bars span the full
radial range of stars in each bin.
Severe crowding causes 
peaks in the RGB, AGB, and total profiles
at $R_{dp} \sim 10 - 14$ arcmin.
The total profile has a steep slope for $R_{dp} \sim 10 - 19$ arcmin
and a shallower slope for larger radii, 
confirming the 
behavior of the diffuse light profile.
The ratio of metal-poor to metal-rich RGB star counts
increases with radius, suggesting a decreasing mean metallicity.
}
\label{fig:sdprofile}
\end{figure}

\begin{figure*}[t]
\epsscale{0.99}
\plotone{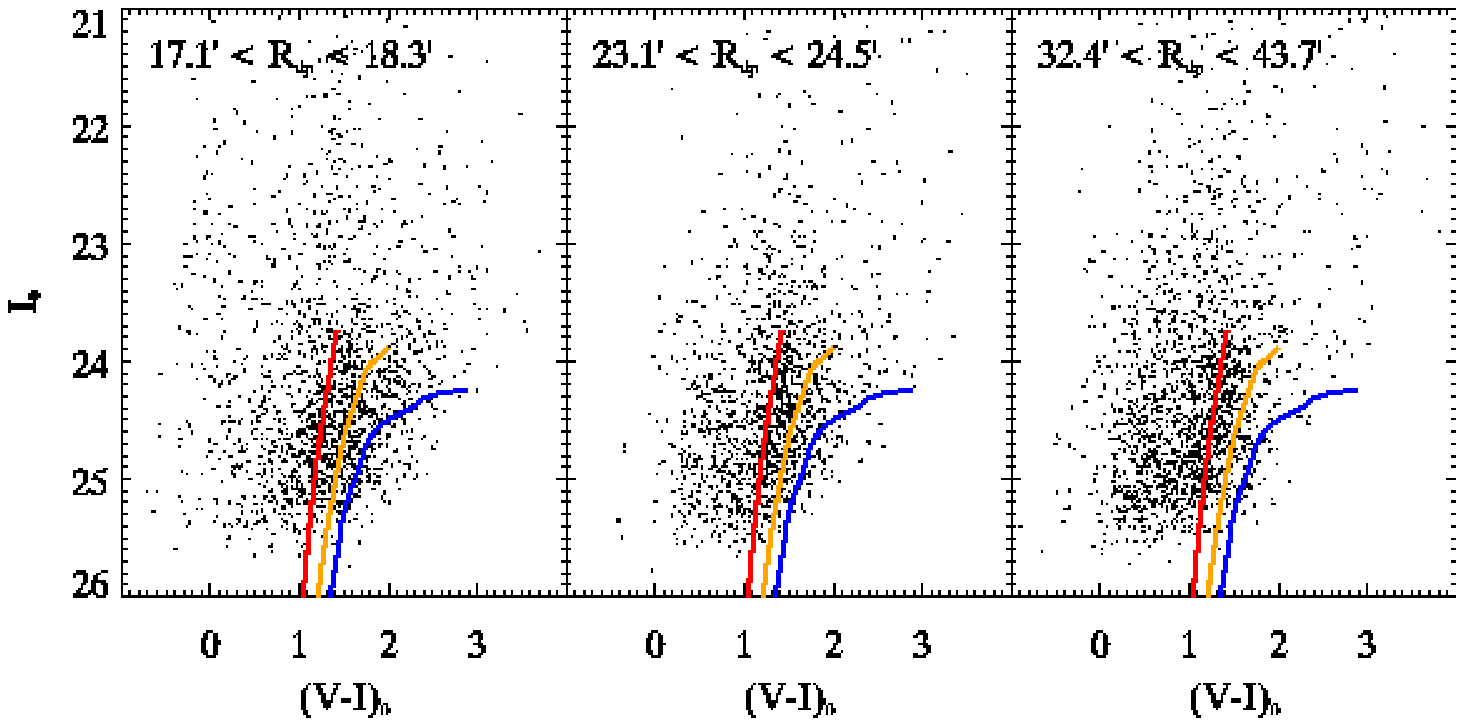}
\caption{Color-magnitude diagrams of point sources in three different
deprojected radial ranges as indicated in the panels.  
Overplotted are theoretical red giant branches
for an age of 10 Gyr and [M/H] = --1.3, --0.7, and --0.4
\citep{Marigo08}.
The diagrams have the same total number of point sources. 
Most of the sources at $I_0 \gtrsim 24$ 
and $0.2 \lesssim (V-I)_0 \lesssim 1.0$
are unresolved background galaxies.
The red giant branch color shifts to the blue
as radius increases, suggesting a decreasing metallicity.
The right panel indicates the extended component has
a peak metallicity in the range $\rm -1.3 \lesssim [M/H] \lesssim -0.7$.}
\label{fig:cmdrbins}
\end{figure*}

In Fig.~\ref{fig:sdprofile}, we show the 
background-subtracted surface density of stars in each
of the CMD selection boxes described in \S \ref{sec:cmd}.
The lines are
color-coded so that the total RGB box is red, 
metal-poor RGB box is orange, metal-rich RGB box is green, 
AGB box is magenta, MS+BHeB box is blue, and RSG box is cyan.
The top profile is the total of all the boxes.
Each point in the profiles is the mean $R_{dp}$
of all stars within a bin.
We chose the bin widths 
so they would have approximately the same number of total stars.
The vertical error bars include Poisson noise and 
systematic uncertainty in the background, while the 
horizontal error bars denote the full radial range of stars in each bin.
To count stars and pixel area, we used only those pixels 
with the highest weights in the confidence maps 
(i.e., with contributions from all dithered images).
We also applied a small crowding correction factor to the profiles, 
typically amounting to $\sim 20\%$, following the
prescription in \citet{Irwin84}. 
This correction factor breaks down 
in the most crowded regions where the stellar PSFs 
significantly overlap, resulting in a turnover 
in the density profiles at $R_{dp} \sim 10 - 14$ arcmin.
Therefore, we limit most of the resolved star analysis to radii beyond 
this turnover.

The young star profiles (MS+BHeB and RSG) exhibit the most substructure, 
as expected from their 2-D spatial distribution.
The broad bump at $\sim 14 - 24$ arcmin is
due to HoIX and the most northern spiral arm.
Beyond 12 arcmin, the vast majority of the young stars
are located in HoIX, Arp's Loop, and 
the northern two spiral arms.
The sharp drop in the young star profiles at 
$R_{dp} \approx 10 - 12$ arcmin approximately coincides 
with the star formation threshold radius in M81
measured by \citet{Martin01} at $R_{HII} \approx 12.3$ arcmin.
They obtained this value from the drop in the 
azimuthally averaged $H\alpha$ surface brightness profile over the
whole disk, but asymmetries cause systematic azimuthal variations in $R_{HII}$, 
of a few arcmin \citep{Martin01}.
Like the young stars, the azimuthally averaged radial profile 
of HI flux shows a peak at
about 8 arcmin, with a pronounced hole near the nucleus
and a steep gradient beyond 8 arcmin, 
so the peak in the young star profiles at this radius may be real
rather than an artifact of incompleteness.

The total star count profile in Fig.~\ref{fig:sdprofile} confirms the
behavior of the diffuse light profile, with a
steep inner slope between 10 and 19 arcmin 
and shallower outer slope beyond 19 arcmin.
The break radius lies 
somewhat farther out than inferred
from the diffuse light profile, but this
difference is not significant given the 
$1\sigma$ uncertainties in the latter.
The fact that this transition to a flatter radial falloff
is seen in both diffuse light and star count analyses means
it is genuine and not due to background uncertainties
or completeness variations.

The RGB stars dominate the total number counts
outside $R_{dp} \sim 12$ arcmin, so the total RGB
profile has a similar flattening as the total star profile.
Interestingly, the metal-poor RGB component has a
shallower slope than the metal-rich component. 
The appearance of such a trend could have a number of causes.
First, the background counts in the metal-poor RGB box would
have to be underestimated by a factor of $\sim 2.5$,
or $\sim 2.0$ if the 
metal-rich RGB box background were simultaneously 
overestimated by a factor of $\sim 10$.
Since these factors are much larger than the differences
between the NGC 4244 and ACS contamination estimates
and their statistical uncertainties, we
consider it unlikely that the observed population gradient
is due to an error in the background subtraction.
It could be conceivable that the
completeness of the metal-rich RGB box has a stronger dependence on
galactocentric radius than the metal-poor RGB box because 
the metal-rich stars are fainter in the V-band.
However, correcting for such a difference would actually enhance the
population gradient by making the metal-rich
profile rise even more sharply to smaller radii.
Another possibility is that the photometric errors could be scattering 
an increasing fraction of stars from the metal-poor RGB
box to the metal-rich box as radius decreases
and the stellar crowding increases.
This also is unlikely because the median errors change by $< 0.05$ mag
for radii $>$ 15 arcmin and this amount 
is much smaller than the widths of the CMD boxes.

The veracity of the population gradient apparent in the RGB density 
profiles is supported by Fig.~\ref{fig:cmdrbins}, which compares
the CMDs for three different radial ranges 
(indicated in the top of each panel), 
moving outwards in radius and starting 
where the two RGB boxes have roughly the same surface density.
The similarity between the faint envelopes of the CMDs
testifies further to the stability of the photometric errors
and completeness rate over this radial range.
Overplotted on the CMDs are theoretical RGBs from \citet{Marigo08} for
[M/H] $=$ --1.3, --0.7, and --0.4 and an age of 10 Gyr.
Most of the sources at $I_0 \gtrsim 24$ 
and $0.2 \lesssim (V-I)_0 \lesssim 1.0$
are unresolved background galaxies.
Each CMD has roughly the same number of objects, so 
the area covered by each and, consequently, 
the number of background galaxies 
progressively increases with radius.

There is a clear excess of sources in all three CMDs where 
we would expect to find an RGB population.
The color distribution of these sources shifts
to the blue as radius increases.
In the innermost CMD, these sources are distributed
roughly evenly between the isochrones, but in the
outermost CMD, the highest density of sources lies  
between the --1.3 and --0.7 isochrones,
indicating a peak $\rm [M/H] \sim -1.0$.
A similar result holds, but shifted downward by $\sim 0.2$ dex, 
if we use the Dartmouth isochrones \citep{Dotter07a} 
with the same age.
Splitting the difference between the stellar models,
we estimate the extended component has a 
peak metallicity $\rm [M/H] \sim -1.1$
in the range $R_{dp} = 32 - 44$ arcmin.
If the RGB stars were as young as 2 Gyr, then their
metallicities would shift upward by $\sim 0.4$ dex.
Because of the age-metallicity degeneracy, we cannot
completely rule out age effects, 
but we consider it unlikely that the observed trend 
is entirely due to an age spread, as no single 
metallicity can span the full color range of the
total RGB selection box.

\begin{figure}
\epsscale{0.95}
\plotone{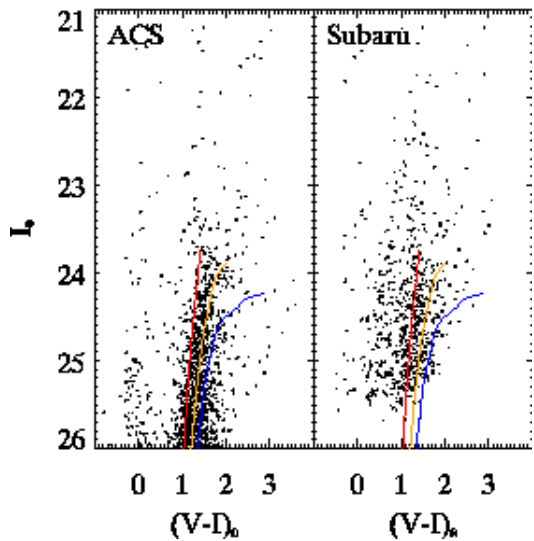}
\caption{
Comparing the color-magnitude diagrams for the
ACS and Subaru data in fields 4 and 5 in Fig.~\ref{fig:fields}
($R_{dp} \sim 20 - 30$ arcmin).
Overplotted are theoretical red giant branches
for an age of 10 Gyr and [M/H] = --1.3, --0.7, and --0.4
\citep{Marigo08}.
The 50\% completeness level in the Subaru CMD
is estimated to lie at $I_0 \sim 25.0$ for
$(V-I)_0 < 1.0$ and $I_0 \sim 24.4$ for $1.0 < (V-I)_0 < 2.0$ 
(see \S \ref{sec:cmd} for details).
The two diagrams 
appear reasonably similar, bearing in mind 
that the Subaru data have a
brighter limiting magnitude and
more background
galaxies, which lie mostly at $I_0 \gtrsim 24$ 
and $0.2 \lesssim (V-I)_0 \lesssim 1.0$.
}
\label{fig:acscmd}
\end{figure}

As discussed in \S \ref{sec:cmd}, 
the Subaru point source catalog contains some unresolved
background galaxies and completeness becomes an important issue
for RGB metallicities [M/H]~$\gtrsim -0.7$,
raising the question of whether background contamination, incompleteness, 
or systematic color-magnitude shifts are 
biasing our metallicity estimate.
We address this question by comparing in Fig.~\ref{fig:acscmd}
the ACS and Subaru CMDs 
for the two ACS background fields. 
The CMDs contain point sources from the
same two regions on the sky (fields 4 and 5 in Fig.~\ref{fig:fields}), 
but the sources are not cross-matched. Thus,
the Subaru CMD contains true stars and unresolved
background galaxies which, as previously discussed, 
are expected to mainly inhabit a region blueward of the RGB. 
The CMDs appear reasonably similar, and there
is no obvious sign in the ACS CMD of any significant 
metal-rich RGB population at this
distance ($R_{dp} \sim 20 - 30$ arcmin)
that could have been missed by the Subaru catalog.
Based on the above considerations, 
we assign an uncertainty of $\pm 0.3$ dex
to our metallicity estimate for the extended component.

Using RGB star counts in several WFPC2 fields across M81, 
\citet{Tikhonov05} presented 
evidence for a flattening in 
stellar surface density at $R_{dp} \sim 20$ arcmin.
Their study, however, relied on only two fields beyond this radius 
on opposite sides of the galaxy
(field 1 in Fig.~\ref{fig:fields} and another coincident with field 3).
From the color of the RGB, they derived [Fe/H] $= -0.6$
in four of their fields at $R_{dp} < 14$ arcmin 
and [Fe/H] $= -0.77$ in 
field 1 of Fig.~\ref{fig:fields}, 
lying at $R_{dp} = 24$ arcmin.
Mouhcine et al. (2005) also analyzed field 1 and
derived [Fe/H] $= -0.9 \pm 0.3$.
The differences
between their metallicity estimates and ours are not surprising
because their estimates come from regions with a dominant or non-negligible
contribution from the bright optical disk, which
is more metal-rich than the extended component.
Moreover, the extended component itself may
have a metallicity gradient, so our
estimate strictly applies to the region
$R_{dp} = 32 - 44$ arcmin.
\section{Discussion}
\label{sec:disc}

\begin{figure}
\epsscale{0.95}
\plotone{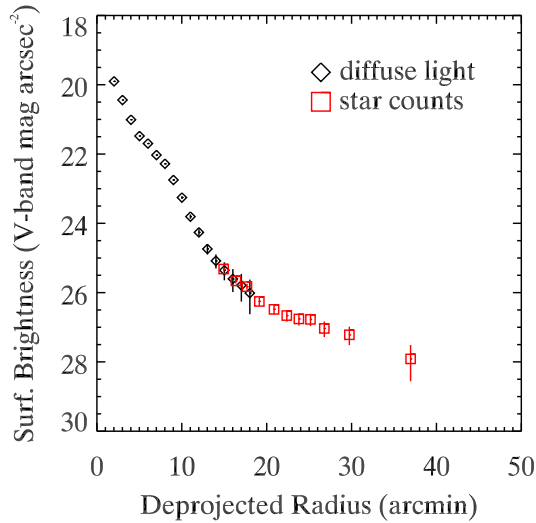}
\caption{Composite V-band surface brightness profile for M81 
created by merging the diffuse light profile (diamonds) 
with the total stellar surface density profile 
(squares).
Errors on the diffuse light are the same as 
in Fig.~\ref{fig:sbprofile}, and errors on the star
counts are from Fig.~\ref{fig:sdprofile}.
}
\label{fig:merged}
\end{figure}

Taking a similar approach as \citet{Irwin05}, 
we construct a composite surface brightness profile
by using the diffuse light and total star count profiles in the
regions where they are each most reliable.
That is, within the bright optical disk, where the effects
of incompleteness are most severe, we use the diffuse light
profile because it is insensitive to these effects.
In the outer regions, where the sky background dominates the
diffuse light, but where the completeness rate is the highest
and varies the least, we use the star counts, which have a higher
contrast over the background than the diffuse light.
The overlapping region allows us to bring the star counts onto the same
absolute luminosity scale as the diffuse light.
By merging the diffuse light and resolved star count profiles
in this way, we can trace M81's surface brightness over a
larger radial range and to fainter magnitudes than is possible
with either profile alone.

The resulting composite V-band light profile (Fig.~\ref{fig:merged})
shows even more clearly the presence of two regimes, a steep
inner gradient for $R_{dp} \lesssim 18$ arcmin and 
a shallower outer gradient at larger radii.
As shown in \S \ref{sec:radial}, the inner
profile gradient follows an exponential with $h = 2.7$ arcmin.
Fitting the total star counts beyond 20 arcmin yields 
$h = 12.9 \pm 0.9$ arcmin and $\gamma = 2.0 \pm 0.2$.
These fits maximize 
the extended component's contribution at small radii because they ignore the
bright optical disk.

\begin{figure}
\epsscale{0.95}
\plotone{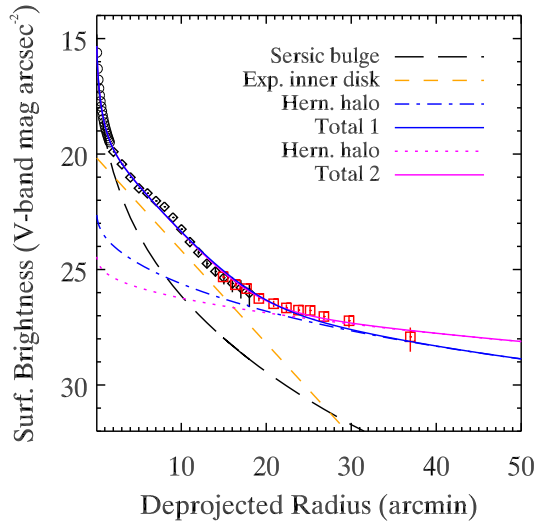}
\caption{
Exploring possible decompositions of M81's
V-band surface brightness profile.
Diffuse light (diamonds) and star counts (squares)
are supplemented with 
the total bulge plus disk model of Mollenhoff (2004) 
(circles) to represent
the bulge region ($R_{dp} \lesssim 2$ arcmin) 
where our images become saturated.
Errors are the same as in Fig.~\ref{fig:merged}.
Broken lines show individual components: 
S\'ersic bulge model of Mollenhoff (2004) 
with $n = 3.0$ and $R_e = 0.75$ arcmin (long dashed), 
exponential inner disk with scale-length 2.7 arcmin (dashed), 
and two spherically symmetric Hernquist halos with scale radii 
of 14.0 arcmin (dot-dashed) and 53.1 arcmin (dotted).
These scale radii are recent estimates for the halos
of the MW \citep{Newberg06} and M31 \citep{Ibata07}, respectively, 
scaled to the distance of M81 (1 arcmin $\approx 1$ kpc).
Two solid lines show the total profiles after adding
the bulge, inner disk, and one Hernquist halo.
}
\label{fig:components}
\end{figure}

In Fig.~\ref{fig:components}, we show 
two possible 3-component decompositions
of the composite light profile.
The point symbols and error bars are the same 
as in Fig.~\ref{fig:merged}.
To represent the bulge region, we have added
the profile of \citet{Mollenhoff04}.
In doing so, we assume his model is a good 
description of the true V-band light distribution.
His total light profile inside 2 arcmin is shown
as circles and his S\'ersic bulge component 
(with $n = 3.0$ and $R_e = 0.75$ arcmin) is the long-dashed line.
The bright optical disk with scale-length 
$h = 2.7$ arcmin is the short-dashed line.
Because of background uncertainties and the limited radial baseline, 
the total profile's shape parameters cannot be very well constrained
in a simultaneous 3-component fit.
Instead, we begin by using the halos of the
MW and M31 as benchmarks for comparison under the
assumption that the extended component is M81's halo.
Later, we consider other possibilities such as a thick disk.

Numerous studies over the years have found that 
the MW stellar halo
has a peak [Fe/H] $\sim -1.6$ (or [M/H] $\sim -1.3$ for
a typical $\rm [\alpha/Fe] \sim 0.4$), 
and a power-law volume density distribution with
exponent $\gamma + 1 \sim 3.0$ 
(see the recent reviews by Helmi [2008] and Geisler et al.\ [2007]).
\citet{Newberg06} also found the distribution of halo F-turnoff stars
in the SDSS followed a Hernquist
profile with best-fit scale radius $r_s \approx 14$ kpc
\footnote{
We recall that $r_s \approx 41\%$ 
of the half-mass radius
and $\approx 55\%$ of the (projected) half-light radius 
of a Hernquist profile \citep{Hernquist90}.}.

Several recent studies have found evidence for an
extended halo in M31 with 
metallicities in the range --0.7 to --1.5
\citep{Reitzel02,Kalirai06,Chapman06,Richardson09}.
In their photometric study of 
the southern quadrant of M31, \citet{Ibata07} showed that
this component could be fit by a Hernquist profile with 
$r_s = 53.1 \pm 3.5$ kpc or an exponential profile
with scale-length $h = 46.8 \pm 5.6$ kpc
out to 150 kpc.
Fitting the minor axis profile in regions devoid of
spatial substructures, they estimated a projected 
power-law exponent of $\gamma = 1.91 \pm 0.12$.

Two Hernquist profiles with the same scale radii as the 
MW and M31 halos translated to M81's distance are represented
by the dot-dashed and dotted lines, respectively, 
in Fig.~\ref{fig:components}.
To plot these profiles, we assumed M81's 
extended component is spherically symmetric
and multiplied the deprojected radius in the appropriate formulae
by a geometrical factor of $\sqrt{0.5\ (1+cos^2(i))} \approx 0.8$,
to approximately account for deprojecting the halos with 
M81's inclination, $i = 58$ deg \citep{Regan94}.
The two Hernquist profiles provide an adequate
description of the data, but are virtually indistinguishable over the
observed radial range.
Extending the survey out to
$R_{dp} \gtrsim 50$ arcmin would offer the best chance
of ruling out one Hernquist model or the other.
Integrating the profiles out to 100 arcmin, 
M81's halo would contain $\sim 10 - 15\%$
of its total V-band luminosity, or 
$L_V \sim 3 - 6 \times 10^9\ L_{\sun}$.
Uncertainties in the diffuse light sky subtraction 
in the range $R_{dp} \sim 14 - 18$ arcmin
translate to a systematic uncertainty of about $\pm 50\%$
in our luminosity estimate, since this was the
region used to normalize the star counts to absolute V-band luminosity.
Nevertheless, M81's extended component appears to be 
more luminous than the MW and M31 halos, 
which have $L_V \sim 10^9\ L_{\sun}$ \citep{Carney90,Ibata07}.

\begin{figure}
\epsscale{0.95}
\plotone{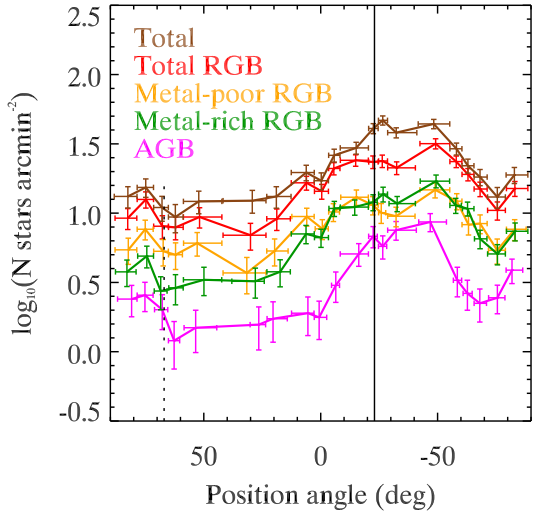}
\caption{Background-subtracted azimuthal variation in stellar surface density
for projected radii of 12 -- 18 arcmin.
Error bars have the same meaning as those in Fig.~\ref{fig:sdprofile}.
The solid and dotted lines mark the major and minor
axes of M81, respectively.
The metal-poor RGB stars have a flatter profile, 
as expected if some of them belong to M81's halo
or a more face-on or thicker disk component.
}
\label{fig:az}
\end{figure}

As we have discussed, the total star counts beyond $R_{dp} \sim 18$
arcmin are dominated by metal-poor RGB stars.
If this metal-poor RGB component is M81's halo, then it should
display a more circular morphology projected on the sky
than the metal-rich RGB component.
We test this possibility in Fig.~\ref{fig:az}, where we examine the
background-subtracted azimuthal variation in 
stellar surface density in a circular annulus 
around M81's nucleus.
To minimize the effects of crowding on the profiles and maximize
the baseline in position angle, we include only stars
with projected radii of 12 -- 18 arcmin.
This range spans deprojected radii of about 15 -- 30 arcmin in total, 
with the precise $R_{dp}$ sampled depending on the position angle.
The solid and dotted vertical lines mark M81's major and minor axes, 
respectively.

The profiles have a roughly sinusoidal shape, as expected for a 
disk morphology with M81's inclination and position angle.
The profile of the metal-poor RGB component appears marginally flatter
than the metal-rich component, suggesting it could contain
a superposition of disk and halo populations, 
or a disk structure with a larger scale height
or lower inclination than the bright optical disk.
Applying a K-S test on the azimuthal distributions of the
metal-poor and metal-rich RGB stars (in the annulus) gives a probability
of 0.1\% that they are drawn from the same parent distribution.
The fact that the maxima of the curves are offset
from the major axis 
is probably caused by the northern 
spiral arms, which host strong star formation 
west of the major axis.

One may wonder whether the most recent close encounters 
with M82 and NGC 3077 
caused any measurable asymmetries in 
the density distribution of M81's halo.
These encounters supposedly occurred $\sim 200 - 300$ Myr ago, which 
is much less than the halo crossing time of 1 Gyr, 
so there should not have been enough time for 
the halo to reach a new equilibrium.
We have examined the stellar density profiles on both sides of M81,
finding similar trends as seen in Fig.~\ref{fig:sdprofile}.
This fact suggests the encounters did not signficantly disturb the
halo, but additional data are needed at other position
angles and reaching larger radii to confirm this suggestion.
A survey of M81's southern region, particularly along
the minor axis, would help further constrain the interaction history
and any role it had in shaping the halo (and disk).

\begin{figure}
\epsscale{0.95}
\plotone{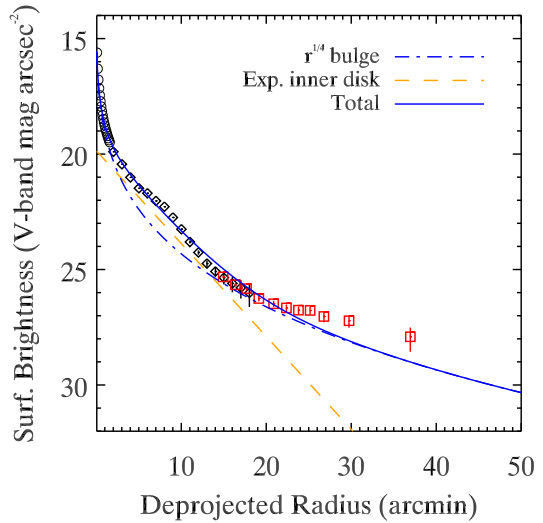}
\caption{Same points as in Fig.~\ref{fig:components}.
The dash-dotted line shows the best-fit $r^{1/4}$ law with
effective radius $R_e = 1.8 \pm 0.1$ arcmin. 
The dashed line is an exponential disk with scale-length
of 2.7 arcmin and the solid line is the total model.
The $r^{1/4}$ law provides a poorer fit than
the Hernquist profiles in Fig.~\ref{fig:components}.
}
\label{fig:sersic}
\end{figure}

Fig.~\ref{fig:sersic} tests the possibility
that the extended component is the faint extension of M81's bulge.
In this figure, the dashed line is the exponential disk with
$h = 2.7$ arcmin, the dot-dashed line is 
our best-fitting $r^{1/4}$ law with effective
radius $R_e = 1.8 \pm 0.1$ arcmin 
and the solid line is the sum of both components.
In this case, the bulge contributes
$\sim 60\%$ of the total V-band luminosity.
The data favor the Hernquist profiles over 
the $r^{1/4}$ law because the latter has a slope
that is too steep at large radii.
The $r^{1/4}$ law would also imply a significant 
metallicity gradient in M81's bulge, 
as \citet{Kong00} estimated [M/H] $\sim 0.2$
inside 2 arcmin.
Fitting a S\'ersic law yields $n = 4.7 \pm 0.3$
and $R_e = 1.7 \pm 0.1$ arcmin, but does not 
significantly improve the fit quality.

Rather than a halo or bulge, M81's extended component 
may be some type of disk structure.
Both metal-poor and metal-rich RGB stars may 
belong to a single perturbed disk component 
with a negative metallicity gradient and a 
change in intrinsic orientation 
and/or scale-length at $R_{dp} \sim 18$ arcmin.  
Alternatively, the extended component may be more analagous to
the MW's thick disk or M31's extended disk \citep{Ibata05}, 
with distinct dynamics and formation history
from the dominant inner disk.
The MW's thick disk has a scale-length roughly $40\%$
larger than the thin disk, or $\sim 3.6$ kpc \citep{Juric08},
it has a mean metallicity 
$\rm \langle [Fe/H]\rangle \sim -0.6$ 
\citep{Gilmore95,Robin96,Soubiran03,AllendePrieto05}, and it
contains $\sim 15\%$ of the total disk light \citep{Buser99,Chen01,Larsen03}.
In comparison, M81's extended component is less centrally concentrated, 
has a slightly lower metallicity, but has a similar luminosity.
We note that the MW's thick disk has been studied mainly 
within a few kpc of the solar cylinder, so it is difficult to 
draw firm conclusions from this comparison.

Regardless of the disk/spheroid nature of the extended component, 
M81 seems to be a nearby example of the
so-called anti-truncated galaxy class identified 
by \citet{Erwin05} and \citet{Pohlen06}.
Some of the anti-truncated envelopes found by \citet{Erwin05}
had rounder isophotes than the inner disks, 
suggesting a transition to a spheroidal component
for those particular galaxies, but
\citet{Pohlen06} could not reliably measure
the outer isophote shapes in their sample.
It is also worth noting that 
the break in M81's profile occurs at
a larger radius 
than nearly all the anti-truncated galaxies in 
\citet{Pohlen06}.

The results presented here may also have important ramifications for studies
of the star formation history (SFH) in M81.
For field 6 in Fig.~\ref{fig:fields}, 
\citet{Williams08} derived a SFH in which $\sim 60 \%$ of stars formed
prior to 8 Gyr ago.
In light of our results, the stellar population in
the \citet{Williams08} field is actually a mixture of the inner disk
and extended component, which at this distance
on the major axis could contribute $\sim 25 - 35\%$
to the total V-band luminosity.
We suggest, therefore, that if the extended component is an
old spheroid or thick disk, then it could be responsible for some of the
star formation in their oldest age bin ($> 8$ Gyr), 
meaning M81's outer disk is younger than 
inferred from the total SFH.
More generally, these considerations highlight the complications that
can arise in pencil-beam studies when  
multiple galactic structural components lie along 
the same line-of-sight.
Studies of the outer disks of massive spirals 
may contain a non-negligible ``contamination'' from their 
spheroids or thick disks and vice-versa.

\section{Summary}
\label{sec:summ}

Using Suprime-Cam on the Subaru telescope, we have
conducted the first wide field mapping of both young 
($\lesssim 200$ Myr) and old ($\sim 1 - 10$ Gyr)
resolved stellar populations around M81, reaching projected distances
of $\sim 30$ kpc.
Throughout the surveyed region, the surface density of 
young stars ($\lesssim 100$ Myr) traces the HI column density
in a manner similar to the Kennicutt-Schmidt relation, 
but we are able to probe the relation over a 
much wider range of gas densities than is 
normally possible with $H\alpha$ imaging.
We find no strong evidence for a population of RGB
stars in Arp's Loop over and above the background M81 population.
This is consistent with the idea that 
Arp's Loop was initially completely gaseous tidal debris that began forming
stars around the time of the last significant gravitational
interaction of the M81 group $\sim 200 - 300$ Myr ago.

Both diffuse light and resolved star counts
show evidence for an extended structural component beyond
the bright optical disk with a
much flatter surface brightness profile, 
but the star counts 
allow us to probe this component to
fainter levels than is possible with the diffuse light alone.
This component begins to dominate 
at $R_{dp} \sim 18$ arcmin and 
$\mu_V \sim 26\ {\rm mag\ arcsec^{-2}}$, 
and continues out to the last measured point at
$R_{dp} \sim 37$ arcmin and 
$\mu_V \sim 28\ {\rm mag\ arcsec^{-2}}$.
This flattening is accompanied
by a shift in RGB color toward the blue, 
which we interpret as a shift to lower metallicity.

The extended component shares some similarities
with the MW's halo and thick disk, but has some
important differences, as well.
If this extended component is $\sim 10$ Gyr old,
then it has a peak metallicity $\rm [M/H] \sim -1.1 \pm 0.3$
at radii $R_{dp} = 32 - 44$ kpc assuming a distance of 3.6 Mpc.
This metallicity is slightly higher than the MW's halo, but
lower than the MW's thick disk.
In the radial and azimuthal range probed by our data, the 
surface density profile of the extended
component follows a power-law with an exponent of
$\gamma \sim 2$, similar to the MW's halo, 
but much shallower than the MW's thick disk.
If it is separate from M81's bulge and thin disk, then it contains 
$\sim 10 - 15\%$ of M81's total V-band luminosity, which
is similar to the MW's thick disk, but several times
more luminous than the MW's halo.
Caution must be exercised when making such comparisons, however, 
since these structures have been isolated with
different selection techniques and probed at
different galactocentric distances.

Other interpretations for the extended component, 
such as a perturbed thin disk, remain viable. 
An interaction induced disturbance occurring 
$\lesssim 1$ Gyr ago seems an unlikely cause for this component, 
as it appears in both the NE and NW
quadrants relative to the galaxy's nucleus.
If this component is the faint extension of the bulge, 
a possibility which our data disfavor 
relative to a halo or thick disk, 
then the bulge would contain over half
of the total V-band luminosity and must have a 
significant metallicity gradient, from super-solar
in the nucleus to about 10 times less than solar at
a projected radius of $\sim 30$ kpc.

Further progress in understanding this structure in M81
requires mapping the resolved RGB stars over a wider area.
This includes extending the radial baseline to properly
test different structural models, and probing more 
position angles to check for asymmetries that may have
occurred as a result of gravitational interactions with its 
galactic neighbors.
Kinematic information for individual
RGB stars in this component would also be important for
constraining its nature, but this must await future 
generations of ground-based telescopes and spectrographs.

\acknowledgements

We thank Rosie Wyse, Jay Gallagher, Peder Norberg, 
Michele Cirasuolo, and Masafumi Yagi for insightful discussions
during the preparation of this paper, and Min Yun for providing
us with his HI map.
MKB and AMNF are supported by a Marie Curie Excellence Grant from
the European Commission under contract MCEXT-CT-2005-025869.
This work is partly supported by a Grant-in-Aid for Science Research
(No.19540245) by the Japanese Ministry of Education, Culture, Sports,
Science and Technology.


\bibliographystyle{apj}
\bibliography{apj-jour,references}




\end{document}